\documentclass[twocolumn,english]{revtex4-2}
\usepackage[T1]{fontenc}
\usepackage[latin9]{inputenc}
\usepackage{color}
\usepackage{amsmath}
\usepackage{graphicx}
\usepackage{comment}
\usepackage{graphicx}
\usepackage{sidecap}
\usepackage[colorlinks=true, allcolors=red]{hyperref}

\makeatletter
\usepackage{babel}

\makeatother

\usepackage{babel}
\begin{document}
\title{Electromagnetic filament coalescence as magnetic island merging with diamagnetic effects}

\author{Souvik Mondal$^{1,3,a}$}
\email{$^{a)}$Email: souvik.mondal@ipr.res.in} 
\author{N Bisai$^{1,3}$}
\author{Abhijit Sen$^{1,3}$}
\author{Indranil Bandyopadhyay$^{1,2,3}$}

\affiliation{$^1$Institute for Plasma Research, Bhat, Gandhinagar 382428, Gujarat, India \\
$^2$ITER-India, Institute for Plasma Research, Bhat, Gandhinagar 382428, Gujarat, India \\
$^3$Homi Bhabha National Institute, Training School Complex, Anushaktinagar, Mumbai 40094, India}

\begin{abstract}
We investigate the nonlinear coalescence of two current-carrying ELM filaments using a three-dimensional electromagnetic fluid model. In the flat-density limit, the coalescence exhibits magnetic island-like reconnection, characterized by X-point formation, current-sheet development, and Sweet-Parker-like resistive scaling. Introducing a blob-like density perturbation modifies the reconnection dynamics: while the peak reconnection rate remains nearly unchanged for weak perturbations, it decreases and is increasingly delayed for larger density amplitudes. Analysis of the induction equation reveals a transition from resistive to increasingly density-dependent advective dynamics. Finite density perturbations also enhance the post-compression rebound, or sloshing, of the filaments. The sloshing amplitude increases with the density-gradient pressure force, establishing density perturbation as an additional control parameter for both reconnection and filament sloshing. These results highlight the coupled electromagnetic and pressure-driven dynamics governing the nonlinear evolution of ELM filaments in the tokamak edge.
\end{abstract}

\maketitle

\section{INTRODUCTION}

Edge-localized modes (ELMs) are transient magnetohydrodynamic instabilities associated with the buildup and sudden release of energy and particles from the pedestal region of magnetically confined plasmas. The resulting intermittent ejection of plasma into the edge and scrape-off-layer (SOL) produces filamentary structures that can contribute significantly to transient particle and heat loads on plasma-facing components \cite{elm_pedestal_2002, H_Zohm_1996_ELM, Banerjee_2021_ELM}. These ELM filaments can exhibit a broad range of sizes, amplitudes, and frequencies, making their nonlinear evolution an important aspect of edge plasma dynamics \cite{Kirk_2006,Wang_2013,Kass_1998}.

Unlike conventional blob structures generated by interchange \cite{Umansky_1998, KRASHENINNIKOV2001368, Zweben_2002, bisai_3f2f_2004, Bisai_2004, Bisai_2005, Shankar_2021} and drift-wave turbulence, ELM filaments can carry a unidirectional parallel current. The existence of such current-carrying ELM filaments has been predicted theoretically and supported by direct experimental measurements \cite{myra_current_carrying_filament_2007,DIIID_ELM_PRL,Current_mes_ELM_PRL,RFX_ELM_PRL}. The unidirectional current produces an electromagnetic interaction between neighboring filaments, providing an additional mechanism for their collective motion \cite{myra_current_carrying_filament_2007,lee_electromagnetic_pop}. This electromagnetic response becomes particularly important in high-$\beta$ edge plasmas, where the inductive magnetic response cannot be neglected \cite{lee_electromagnetic_2015,stepanenko_impact_2020}.

Neighboring current-carrying structures can therefore attract and coalesce through electromagnetic interaction and magnetic reconnection. For ELM filaments carrying parallel currents, this attraction can drive poloidal motion and facilitate merging, while curvature effects can promote radial transport \cite{myra_current_carrying_filament_2007,souvik_pop}. The competition between these effects determines the subsequent filament motion and can modify the structure and transport properties of the resulting coherent plasma structure.

Interacting ELM filaments can also exhibit rotational and oscillatory dynamics. Their density structure generates a self-consistent electrostatic potential and associated $E\times B$ flow, allowing filament rotation \cite{elm_rotation_expt_NSTX}. Previous numerical studies have also shown oscillatory or \textit{sloshing} behavior in the separation of interacting current-carrying filaments \cite{souvik_pop}. The merging dynamics of such current-carrying ELM filaments have further been investigated in the warm-ion regime, where finite ion temperature introduces additional filament deformation and modifies the merging dynamics \cite{souvik_pop_ion}. Similar rebound dynamics occur during magnetic-island coalescence, where current-sheet compression is followed by reconnection and relaxation \cite{D_A_Knoll_Pop_2006, D_A_Knoll_PRL_2006}. These similarities motivate the question of whether electromagnetic filament coalescence can be understood fundamentally as a magnetic-reconnection process.

Magnetic reconnection is a fundamental plasma process through which magnetic energy is converted into plasma kinetic and thermal energy while magnetic topology changes. In collisional plasmas, the classical Sweet--Parker model describes resistive reconnection through the formation of a thin current sheet and the balance between plasma inflow and resistive diffusion \cite{Sweet_1958,parker_1957}. A closely related process occurs during magnetic-island coalescence, where neighboring current-carrying structures approach, compress the magnetic field between them, and form a reconnecting current sheet \cite{Coalescence_instability_finn_kaw,D_A_Knoll_Pop_2006,Yamada_2010}. However, whether the electromagnetic coalescence of filamentary structures in the tokamak edge can be quantitatively connected to classical magnetic-island merging has not been established.

ELM filaments are finite-amplitude density perturbations rather than purely magnetic structures. Their density structure couples the electromagnetic evolution to plasma pressure and introduces additional $E\times B$ and diamagnetic dynamics. Consequently, the interaction between neighboring current-carrying filaments can be governed not only by their mutual electromagnetic attraction and resistive magnetic diffusion, but also by pressure-gradient forces and plasma flows. These effects provide an important physical framework for examining how realistic density structure modifies filament interaction, magnetic reconnection, and the subsequent motion of current-carrying ELM filaments.

In this work, we demonstrate that, in the absence of a significant pressure gradient, the coalescence of two current-carrying filaments exhibits a direct correspondence with magnetic-island merging. The interaction produces a compressed current layer, followed by X-point formation and magnetic reconnection, with the reconnection dynamics showing characteristics consistent with the classical resistive picture. In particular, the characteristic reconnection scaling is consistent with the Sweet-Parker framework. This establishes a direct connection between electromagnetic filament coalescence and magnetic-island merging and shows that filament coalescence can be viewed as a reconnection-mediated process.

We then introduce a blob-like density perturbation to investigate how finite pressure gradients modify this correspondence. The density perturbation introduces pressure-gradient and diamagnetic effects that alter the dynamics of the reconnection layer. Diamagnetic drifts can modify the effective plasma inflow relative to the magnetic flux, leading to asymmetric current-sheet evolution and a reduced reconnection response \cite{Swisdak_2003}. Consequently, classical resistive scaling is no longer sufficient to describe the reconnection dynamics in the finite-density case.


This paper is organized as follows. Section II presents the normalized three-dimensional electromagnetic fluid model. Section III describes the numerical setup and initial and boundary conditions. Section IV presents the results, focusing on filament coalescence, magnetic reconnection, and density-perturbation-induced sloshing. Finally, Section V summarizes the main findings and their implications for the nonlinear dynamics of ELM filaments in the edge regions.

\section{MODEL EQUATIONS\label{sec:Model_Equations}}

The nonlinear evolution of current-carrying plasma filaments is investigated using a reduced electromagnetic two-fluid model derived from the Braginskii fluid equations~\cite{braginskii1965transport}. The model describes low-frequency plasma dynamics in the edge and scrape-off layer (SOL) of a tokamak, where the characteristic frequencies are much smaller than the ion gyrofrequency, and the perpendicular length scales are comparable to the ion sound gyroradius ($\rho_s=c_s/\Omega_s$). A local Cartesian slab geometry is adopted with the equilibrium magnetic field directed along the $z$-axis, while the $x$ and $y$ directions correspond to the radial and poloidal directions, respectively.

In the present work, the ions are assumed to be cold ($T_i=0$), whereas the electrons are treated as an isothermal fluid with a constant electron temperature ($T_e=\mathrm{constant}$). Under these assumptions, the ion pressure contribution is neglected, and the plasma dynamics are governed by the interplay between the electron pressure gradient, magnetic curvature, and electromagnetic forces. The plasma is assumed to remain quasi-neutral, and electron inertia is neglected since the electron-ion collision time is much shorter than the characteristic time scale of filament evolution \cite{stepanenko_impact_2020}.

Unlike electrostatic blob models, the present formulation explicitly includes the inductive parallel electric field,
\begin{equation}
E_{\parallel}^{\rm ind}
=
-\frac{1}{c}\frac{\partial A_{\parallel}}{\partial t},
\end{equation}
which becomes important in high-$\beta$ plasmas where magnetic fluctuations are no longer negligible. Consequently, the model is capable of describing the interaction and coalescence of unidirectional current-carrying filaments through their self-generated magnetic fields.

The governing equations are obtained from particle conservation, current continuity, the parallel electron momentum equation (generalized Ohm's law), and Ampere's law. The dimensional equations can be written as~\cite{lee_electromagnetic_2015,stepanenko_impact_2020,souvik_pop}

\begin{equation}
\frac{e\rho_{s}^{2}}{T_{e}}n\frac{d\omega}{dt}=\frac{1}{e}\nabla_{\parallel}J_{\parallel}-\frac{g_i}{\Omega_s}\frac{\partial n}{\partial y},
\label{eq:unnormalized_vorticity}
\end{equation}

\begin{equation}
\frac{dn}{dt}=\frac{1}{e}\nabla_{\parallel}J_{\parallel}-\frac{g_i}{\Omega_s}
\frac{\partial n}{\partial y},\label{eq:unnormalized_continuity}
\end{equation}

\begin{equation}
-\frac{e}{m_ec}\frac{dA_{\parallel}}{dt}=\frac{e}{m_e}\frac{\partial\phi}{\partial z}-\frac{T_e}{m_e}\nabla_{\parallel}\ln n+\frac{e}{m_e\sigma_{\parallel}}J_{\parallel},
\label{eq:vector_potential}
\end{equation}

where $n$, $\phi$, $\omega=\nabla_\perp^2\phi$, $J_\parallel$, and $A_\parallel$ denote the plasma density, electrostatic potential, vorticity, parallel current density, and parallel magnetic vector potential, respectively. The ion sound gyroradius is defined as $\rho_s=c_s/\Omega_s$, where $c_s=\sqrt{T_e/m_i}$ is the ion sound speed and $\Omega_s=eB_0/m_ic$ is the ion gyrofrequency. The curvature drive is represented by $g_i=2c_s^2/R$, with $R$ being the tokamak major radius.

The parallel current density is related to the electron and ion parallel velocities through $J_{\parallel}=ne(V_{i\parallel}-V_{e\parallel})$, while the Spitzer conductivity is $\sigma_{\parallel}={1.96n_0e^2}/{m_e\nu_{ei}}$, where the electron-ion collision frequency is $\nu_{ei}=2.9\times10^{-6}{n_0\ln\Lambda}/{T_e^{3/2}}$, with $\ln\Lambda\simeq10$.

The convective derivative describing perpendicular $E\times B$ transport is
\begin{equation}
\frac{d}{dt}=\frac{\partial}{\partial t}+\frac{c}{B_0}\hat{\mathbf b}_0\times \nabla\phi \cdot \nabla,
\end{equation}

whereas the magnetic-field-aligned derivative is
\begin{equation}
\nabla_{\parallel}=\frac{\partial}{\partial z}+\frac{\nabla A_{\parallel}}{B_0}\times \hat{\mathbf b}_0\cdot \nabla.
\end{equation}

The parallel current is coupled to the magnetic vector potential through Amp\`ere's law,
\begin{equation}
J_{\parallel}=-\frac{c}{4\pi}\nabla_\perp^2A_{\parallel}.
\end{equation}

Furthermore, since the emphasis of this work is on the perpendicular interaction of current filaments, the equations are averaged along the magnetic field line. Under this approximation, the parallel operators become

\begin{equation}
\nabla_{\parallel}J_{\parallel}=\frac{1}{B_0}[J_{\parallel},A_{\parallel}]_{x,y},
\end{equation}

\begin{equation}
\nabla_{\parallel}\ln n=\frac{1}{B_0}[\ln n,A_{\parallel}]_{x,y},
\end{equation}

where
$[f,g]=(\partial_xf)(\partial_yg)-(\partial_yf)(\partial_xg)$
denotes the Poisson bracket.

To numerically simulate 
Eqs.~(\ref{eq:unnormalized_vorticity})-(\ref{eq:vector_potential}), we use the following normalizations:
${n}/{n_{0}}=\hat{n}$, $t\Omega_{s}=\hat{t}$, ${(x,y)}/{\rho_{s}}=(\hat{x},\hat{y})$, $v_{\parallel}/{c_{s}}=\hat{v}_{\parallel}$, ${J_{\parallel}}/{n_{0}ec_{s}}=\hat{J}_{\parallel}$, ${e\phi}/{T_{e0}}=\hat{\phi}$, ${A}/{B_{0}\rho_{s}}=\hat{A}$, ${T_{e}}/{T_{e0}}=\hat{T}_e$ and $\rho_s{\nabla_{\parallel}}=\hat{\nabla}_{\parallel}$ where $n_0$, $T_{e0}$, and $B_0$ represent the values of plasma density, electron temperature, and toroidal magnetic field in the Last Closed Flux Surface (LCFS). $v_\|$ represents the parallel velocity of the electrons. After normalization (the hats being omitted for simplicity), the reduced model becomes

\begin{equation}
\frac{\partial n}{\partial t}=-[\phi,n]+\frac{\partial J_\parallel}{\partial z}-[A_\parallel,J_\parallel]-g\frac{\partial n}{\partial y},
\label{eq:norm_density_continuty_eq}
\end{equation}

\begin{equation}
n\frac{\partial\omega}{\partial t}=-n[\phi,\omega]+\frac{\partial J_\parallel}{\partial z}-[A_\parallel,J_\parallel]-g\frac{\partial n}{\partial y},
\label{eq:norm_vorticity_eq}
\end{equation}

\begin{equation}
\frac{\partial A_\parallel}{\partial t}=\frac{\partial}{\partial z}(\ln n-\phi)-[A_\parallel,\ln n]+\eta\nabla_\perp^2A_\parallel,
\label{eq:norm_vector_pot_eq}
\end{equation}

together with

\begin{equation}
J_\parallel=-a\nabla_\perp^2A_\parallel,
\label{eq:norm_maxwell_eq}
\end{equation}

where $\omega=\nabla_\perp^2\phi$ is the normalized vorticity, $g=\rho_s/R$ represents the normalized magnetic curvature, $\eta=1/(\Omega_s\tau_s)$ is the normalized resistivity with $\tau_s=4\pi\sigma_\parallel\rho_s^2/c^2$ denoting the magnetic diffusion time, and $a=1.96m_i/(m_e\nu_{ei}\tau_s)$ is the normalized coefficient relating the parallel current density to the magnetic vector potential.

It is worth emphasizing that the present model naturally contains two physically distinct limits. When the density perturbation is absent ($n_b=0$), the pressure-gradient terms vanish, and the dynamics are governed solely by the electromagnetic interaction between parallel current filaments, providing a direct analogue of the classical magnetic island coalescence problem. In contrast, blob-like finite density perturbations ($n_b>0$) introduce pressure-driven effects through the density evolution and induction equations, allowing us to investigate how plasma pressure modifies current-sheet formation, magnetic reconnection, and the subsequent sloshing dynamics.


\section{Numerical Simulation}
\label{sec:Numerical_simulation}


The numerical simulations are carried out by solving Eqs.~(\ref{eq:norm_density_continuty_eq})--(\ref{eq:norm_maxwell_eq}) using the BOUT++ framework under plasma conditions representative of the edge region of a high-$\beta$ tokamak. The reference plasma parameters employed throughout this work are listed in Table~\ref{table:1}. These parameters correspond to typical conditions near the last closed flux surface (LCFS) and are consistent with ITER-relevant edge plasmas reported in previous electromagnetic filament studies~\cite{lee_electromagnetic_pop}. Unless otherwise specified, all variables presented in this paper are normalized according to the normalization introduced in Sec.~II.

\begin{table}
\centering
\setlength{\tabcolsep}{12pt}
\begin{tabular}{l c c} 
 \hline
 Parameter & Value & Unit \\ [0.5ex] 
 \hline\hline
 $n_0$ & $1\times10^{14}$ & cm$^{-3}$ \\ 
 $T_e$ & 200 & eV \\
 $B_0$ & $5.3\times 10^4$ & G \\
 $R$ & 600 & cm \\
 $L$ & $10^4$ & cm \\
 $c_s$ & $9.98\times10^{6}$ & cm/s \\ 
 $\Omega_s$ & $2.64\times10^{8}$ & s$^{-1}$ \\
 $\rho_s$ & $3.78\times10^{-2}$ & cm \\
 $\nu_{ei}$ & $1.02\times10^{6}$ & s$^{-1}$ \\
 $\sigma_{\|}$ & $4.82\times10^{16}$ & s$^{-1}$ \\
 $g_i$ & $3.3\times 10^{11}$ & cm/s$^{2}$\\
 $g$ & $1.25\times 10^{-4}$ & normalized \\
 $\delta$ & 0.7 & cm \\ [1ex] 
 \hline
\end{tabular}
\caption{Typical high-$\beta$ plasma parameters relevant to ITER-like edge conditions. Plasma parameters near the LCFS are used in the simulations~\cite{lee_electromagnetic_pop}.}
\label{table:1}
\end{table}

The computational domain has dimensions $L_x=L_y=256\rho_s$ in the perpendicular plane and $L_z=256384\rho_s$ along the magnetic field direction. Accordingly, two sets of simulations are performed. In the first set, the density perturbation is removed ($n_b=0$) to investigate the flat-density limit and establish its correspondence with the classical magnetic island coalescence problem. In the second set, finite density perturbations ($n_b>0$) are introduced to examine how plasma pressure modifies the reconnection, energy conversion, and sloshing dynamics during current filament coalescence.

The initial plasma density is prescribed as the superposition of two Gaussian perturbations centered at the same radial position but separated in the poloidal direction,
\begin{equation}
\begin{split}
n(\textbf{r},0) = 1 + n_b \exp\left[-\frac{(x-x_0)^2}{\delta^2}\right]\cdot 
\left[\exp\left(-\frac{(y-y_1)^2}{\delta^2}\right) \right.\\
\left. + \exp\left(-\frac{(y-y_2)^2}{\delta^2}\right)\right]
\end{split}
\end{equation}
where $x_0=L_x/2$ denotes the initial radial location of the filament centers, while $y_1=0.325L_y$ and $y_2=0.675L_y$ specify their initial poloidal positions. The parameter $n_b$ controls the amplitude of the density perturbation. In the present work, $n_b$ is treated as a control parameter and is varied systematically to examine how finite density perturbations modify the reconnection and coalescence dynamics. In the absence of a filament-like density perturbation ($n_b=0$), referred to hereafter as the ``flat-density-perturbation'' case, the current-carrying filaments evolve without density-gradient-driven interaction with the surrounding background plasma, whereas finite values of $n_b$ represent blob-like density structures.

The perpendicular size of each filament is characterized by the parameter $\delta$, chosen according to the inertial blob scaling,
\begin{equation}
\delta=\rho_s
\left(
\frac{g_iL^2}
{4c_s^2\rho_s}
\right)^{1/5},
\end{equation}
which yields $\delta\simeq0.7$~cm for the parameters listed in Table~\ref{table:1}. This value is comparable to experimentally measured blob sizes in the edge region of tokamak plasmas~\cite{dippolito_convective_2011}.

The equilibrium parallel current density is initialized with the same spatial profile as the density perturbation,
\begin{equation}
\begin{split}
J_\|(\textbf{r},0) = J_0 \exp\left[-\frac{(x-x_0)^2}{\delta^2}\right]\cdot 
\left[\exp\left(-\frac{(y-y_1)^2}{\delta^2}\right) \right.\\
\left. + \exp\left(-\frac{(y-y_2)^2}{\delta^2}\right)\right]
\end{split}
\end{equation}
where $J_0$ represents the peak parallel current density. Since both filaments carry current in the same direction, they interact through the attractive Lorentz force, providing a suitable configuration for investigating electromagnetic filament coalescence.

The initial parallel magnetic vector potential, $A_\|$, is obtained self-consistently by solving the perpendicular Amp\`ere equation [Eq.~(\ref{eq:norm_maxwell_eq})] using a numerical Laplace inversion. The electrostatic potential and vorticity are initially set to zero, i.e., $\phi=0$ and $\nabla_\perp^2\phi=0$, allowing the plasma flow to develop self-consistently from the electromagnetic evolution.

Periodic boundary conditions are imposed in the poloidal ($y$) direction, whereas homogeneous Neumann boundary conditions are applied in both the radial ($x$) and parallel ($z$) directions for all evolved variables, namely $n$, $\phi$, $\nabla_\perp^2\phi$, $J_\|$, and $A_\|$.

The governing equations are solved using the BOUT++ simulation framework~\cite{DUDSON_bout++}. Fourth-order central finite-difference schemes are employed for spatial derivatives in the radial and parallel directions, while Fourier methods are used along the periodic poloidal direction. The nonlinear advective terms are discretized using a third-order weighted essentially non-oscillatory (WENO) upwind scheme, and temporal integration is carried out with the CVODE implicit solver. Unless stated otherwise, all simulations use a computational grid of $516\times512\times32$ points with uniform grid spacings of $dx=dy=0.5\rho_s$ and $dz=8012\rho_s$, which provide numerically converged solutions and good conservation of the total energy \cite{souvik_pop}.


\section{Simulation results}


In this section, we present numerical simulations of ELM-like current-carrying filaments using the normalized three-dimensional fluid model described in Sec.~\ref{sec:Model_Equations}. Our primary objective is to examine the physical connection between the coalescence of current-carrying filaments and the classical magnetic-island coalescence process. We first consider the flat-density limit, in which the absence of a filament-like density perturbation allows the electromagnetic interaction and reconnection dynamics to be investigated without the influence of pressure-gradient effects. The merging process is then examined over a range of resistivities to characterize the role of resistive dissipation in the filament coalescence dynamics. Finally, we introduce a low-amplitude filament-like density perturbation to investigate how finite density and pressure gradients modify the reconnection process and the subsequent filament merging dynamics.


\subsection{Current Filament Coalescence in the Flat-density Limit}

In this section, we demonstrate that the merging of two current-carrying filaments in the flat-density limit is physically equivalent to the classical magnetic island coalescence problem. We first examine the evolution of the magnetic topology and current sheet formation, followed by the corresponding density evolution. The remaining diagnostics, including the magnetic perturbation, reconnection rate, energy conversion, resistivity scaling, and filament separation, are subsequently employed to establish both the qualitative and quantitative correspondence between the two systems.

\begin{figure*}
    \centering
    \includegraphics[width=0.9\linewidth]{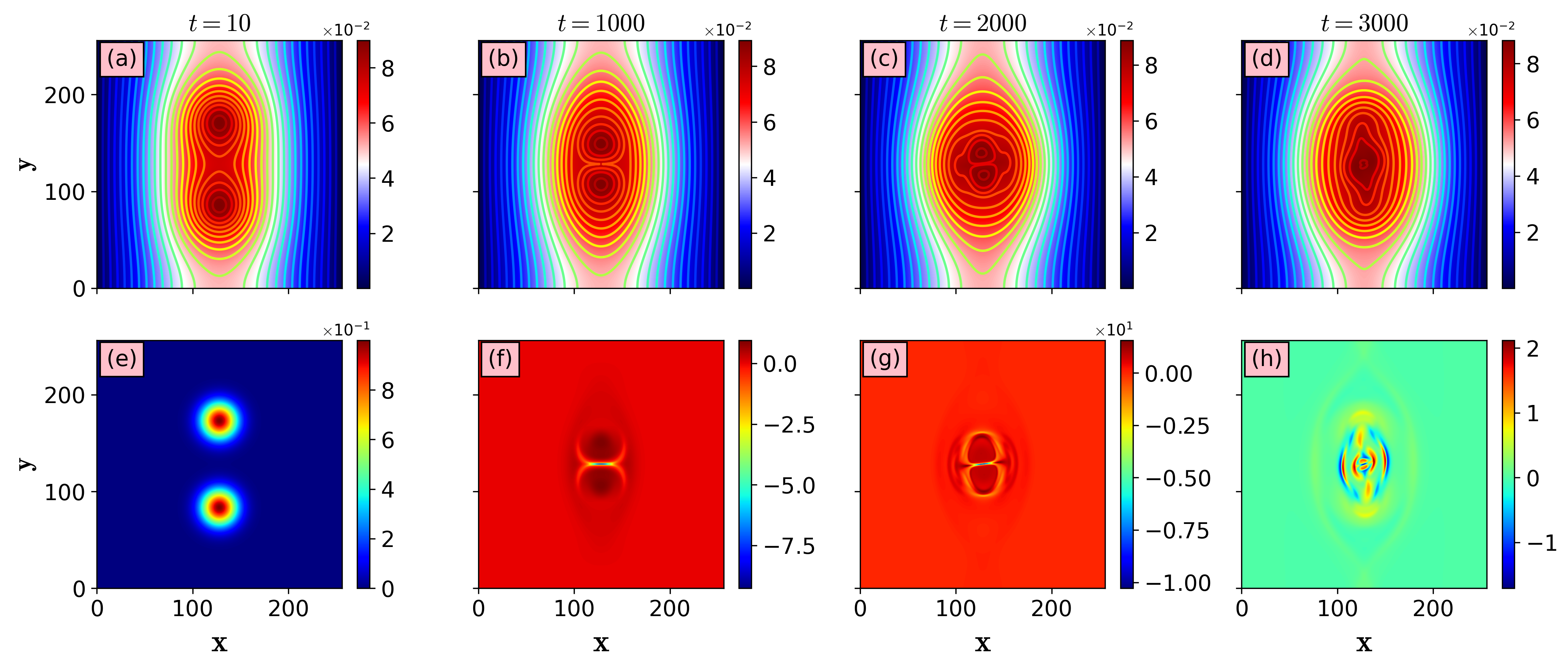}
    \caption{Temporal evolution of the magnetic vector potential $A_{\parallel}$ (top row), shown by color and contour, and parallel current density $J_{\parallel}$ (bottom row) during the coalescence of two current-carrying filaments for the flat-density case ($n_b=0$). The distributions are shown at (a,e) $t=10$, (b,f) $t=1000$, (c,g) $t=2000$, and (d,h) $t=3000\,\Omega_s^{-1}$. The initially separated filaments approach each other and progressively develop a compressed current layer between them, followed by magnetic reconnection and the formation of a single merged magnetic structure at later times.}
    \label{fig:ap_jp_uni}
\end{figure*}

We first examine the evolution of the magnetic topology during the interaction of two unidirectional current-carrying filaments. Figure~\ref{fig:ap_jp_uni} shows the time evolution of the parallel magnetic vector potential, $A_\|$, in the upper panels (a--d) and the corresponding parallel current density, $J_\|$, in the lower panels (e--h). The initial configuration consists of two identical bi-Gaussian current filaments carrying a peak current density of $J_0=1.6~\mathrm{MA/m^2}$ and separated by $90\,\rho_s$ along the poloidal ($y$) direction, as described in Sec.~\ref{sec:Numerical_simulation}. The corresponding magnetic vector potential is obtained self-consistently from Ampere's law,
\begin{equation}
\nabla_\perp^2A_\|=-J_\|.
\end{equation}

At $t=10~\Omega_s^{-1}$, two well-separated magnetic structures are formed around the current channels. Since the currents flow in the same direction, the attractive Lorentz force drives the filaments toward each other, causing compression of the magnetic flux surfaces in the interaction region. By $t=1000~\Omega_s^{-1}$, the separation between the filaments is considerably reduced, and a thin current sheet develops at the midpoint between them, indicating the onset of magnetic reconnection. Simultaneously, the current density becomes strongly concentrated within the reconnection layer owing to the increasing magnetic field gradients.

As the evolution proceeds to $t=2000~\Omega_s^{-1}$, the current sheet becomes thinner and more intense, while the magnetic topology is significantly modified through reconnection. The initially distinct magnetic structures begin to merge into a common flux system, marking the nonlinear stage of the coalescence process. During this phase, magnetic energy stored in the current filaments is released and subsequently converted into plasma kinetic energy.

At $t=3000~\Omega_s^{-1}$, the coalescence process is nearly complete. The two magnetic structures merge into a single island-like configuration, and the current sheet broadens as the reconnection layer relaxes. The entire sequence of magnetic flux compression, current-sheet formation, magnetic reconnection, and final merger closely resembles the nonlinear evolution reported in the classical magnetic island coalescence problem. Therefore, in the absence of density gradients, curvature, and diamagnetic effects, the interaction of current-carrying filaments is governed predominantly by reconnection physics and may be regarded as analogous to the magnetic island coalescence process.

\begin{figure}
    \centering
    \includegraphics[width=0.99\linewidth]{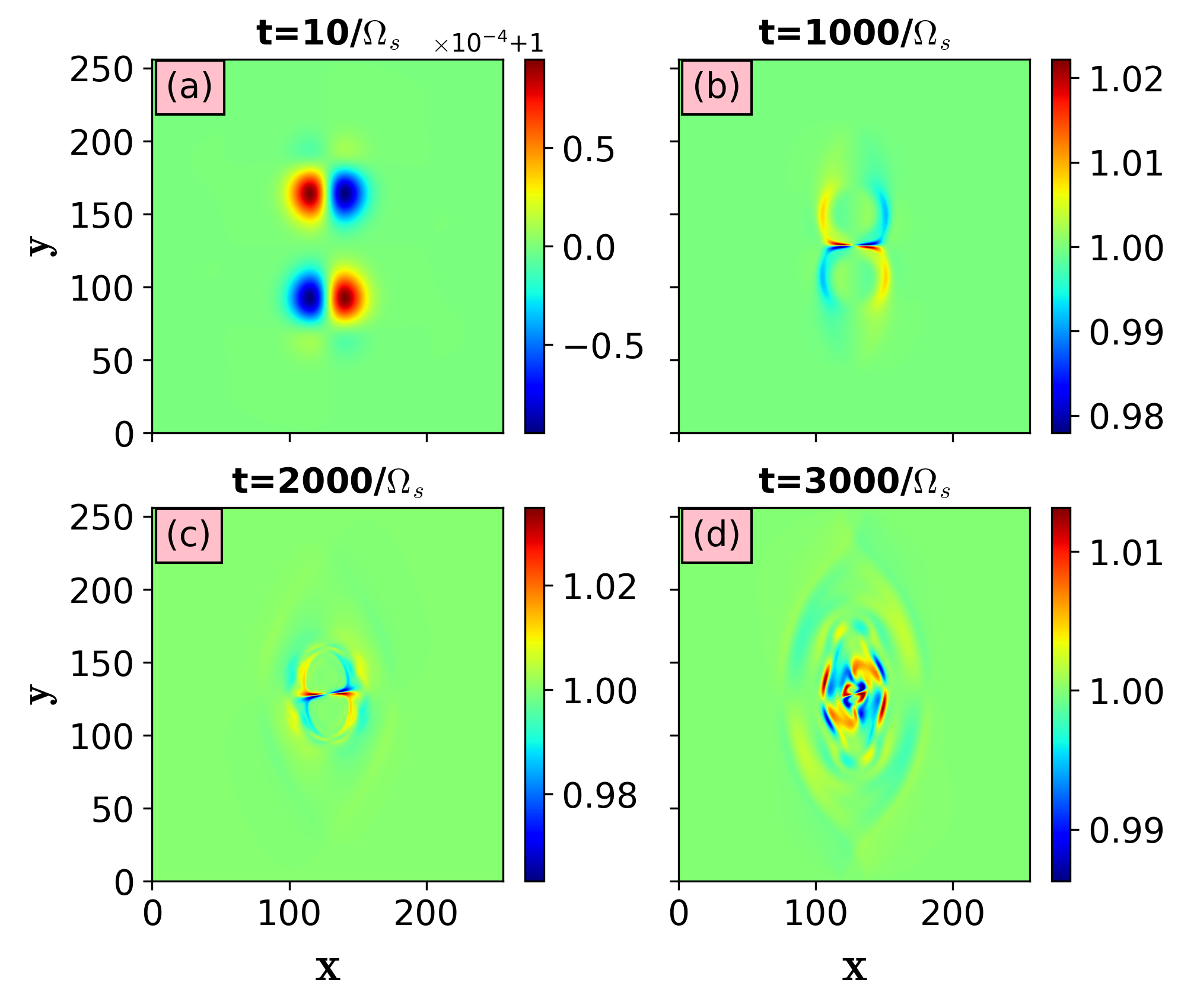}
    \caption{Temporal evolution of the density perturbation for the flat-density case with an initially imposed small perturbation, shown at (a) $t=10$, (b) $t=1000$, (c) $t=2000$, and (d) $t=3000\,\Omega_s^{-1}$. The initially localized density structures evolve as the two filaments approach and interact, followed by the development of increasingly distorted and oscillatory density structures during the later stages of the coalescence.}
    \label{fig:n_uni}
\end{figure}

To verify the validity of the flat-density approximation, we next examine the corresponding plasma density evolution. Figure~\ref{fig:n_uni} shows the density evolution for the case in which no initial density perturbation is imposed ($n_0=1$ and $n_b=0$). Throughout the simulation, the density remains nearly uniform, with only weak perturbations ($\delta n/n_0\ll1$) generated self-consistently through the electromagnetic interaction of the current filaments. As the current sheet develops and reconnection proceeds, localized density perturbations appear around the interaction region due to the induced $E\times B$ flows; however, their amplitude remains negligible compared with the background density. The absence of appreciable density evolution confirms that pressure-gradient and diamagnetic effects play an insignificant role in the present simulations. Consequently, the filament dynamics are governed almost entirely by electromagnetic interactions and magnetic reconnection, making the flat-density limit an ideal framework for establishing the correspondence between current filament merging and the classical magnetic island coalescence problem.

\begin{figure}
    \centering
    \includegraphics[width=0.99\linewidth]{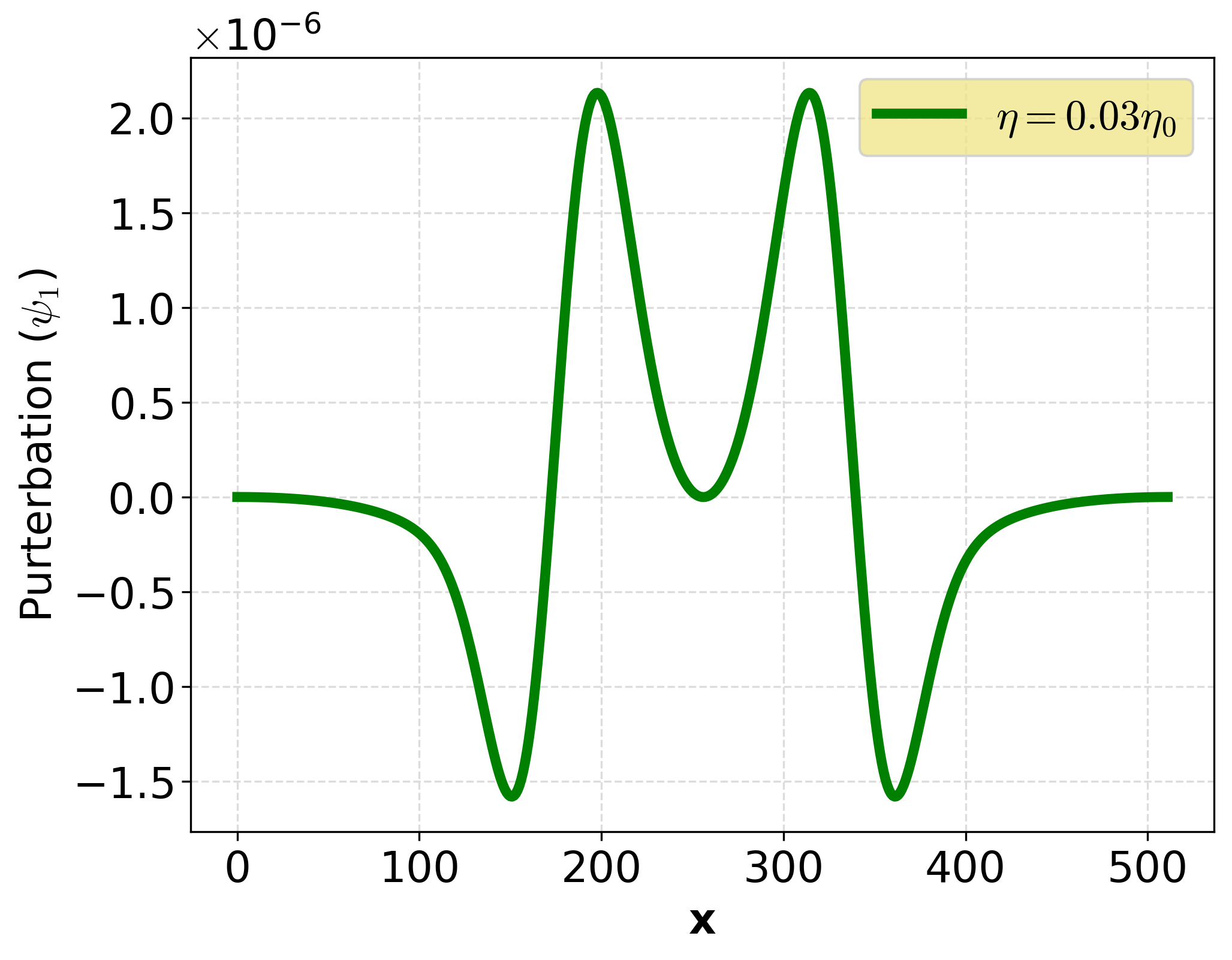}
    \caption{Radial profile of the magnetic vector potential perturbation $\psi_1$ during the linear stage for $\eta=0.03\eta_0$.}
    \label{fig:psi_uni}
\end{figure}

Having established that the density remains nearly uniform throughout the evolution, we next compare the initial magnetic perturbation with the classical eigenmode of the magnetic island coalescence instability. Figure~\ref{fig:psi_uni} shows the radial profile of the magnetic vector potential perturbation, $\psi_1$, during the linear stage of the evolution. The perturbation is obtained directly from the simulated magnetic vector potential by subtracting the equilibrium component, i.e.,
\begin{equation}
\psi_1=A_\|-A_{\|0}.
\end{equation}
The resulting perturbation exhibits a symmetric quadrupolar structure with alternating positive and negative extrema about the filament centers. This profile is qualitatively identical to the perturbation reported by Pritchett \textit{et al.}~\cite{Pritchett_1979}, which was obtained analytically using the linear eigenfunction derived by Finn and Kaw~\cite{Coalescence_instability_finn_kaw} for the magnetic island coalescence instability. The close agreement between the simulated perturbation and the classical linear eigenmode demonstrates that the present current-filament system evolves through the same linear coalescence mechanism. This constitutes the first direct evidence that, in the flat-density limit, the merging of current-carrying filaments is governed by the same linear physics as the magnetic island coalescence problem.

\begin{figure}
    \centering
    \includegraphics[width=0.99\linewidth]{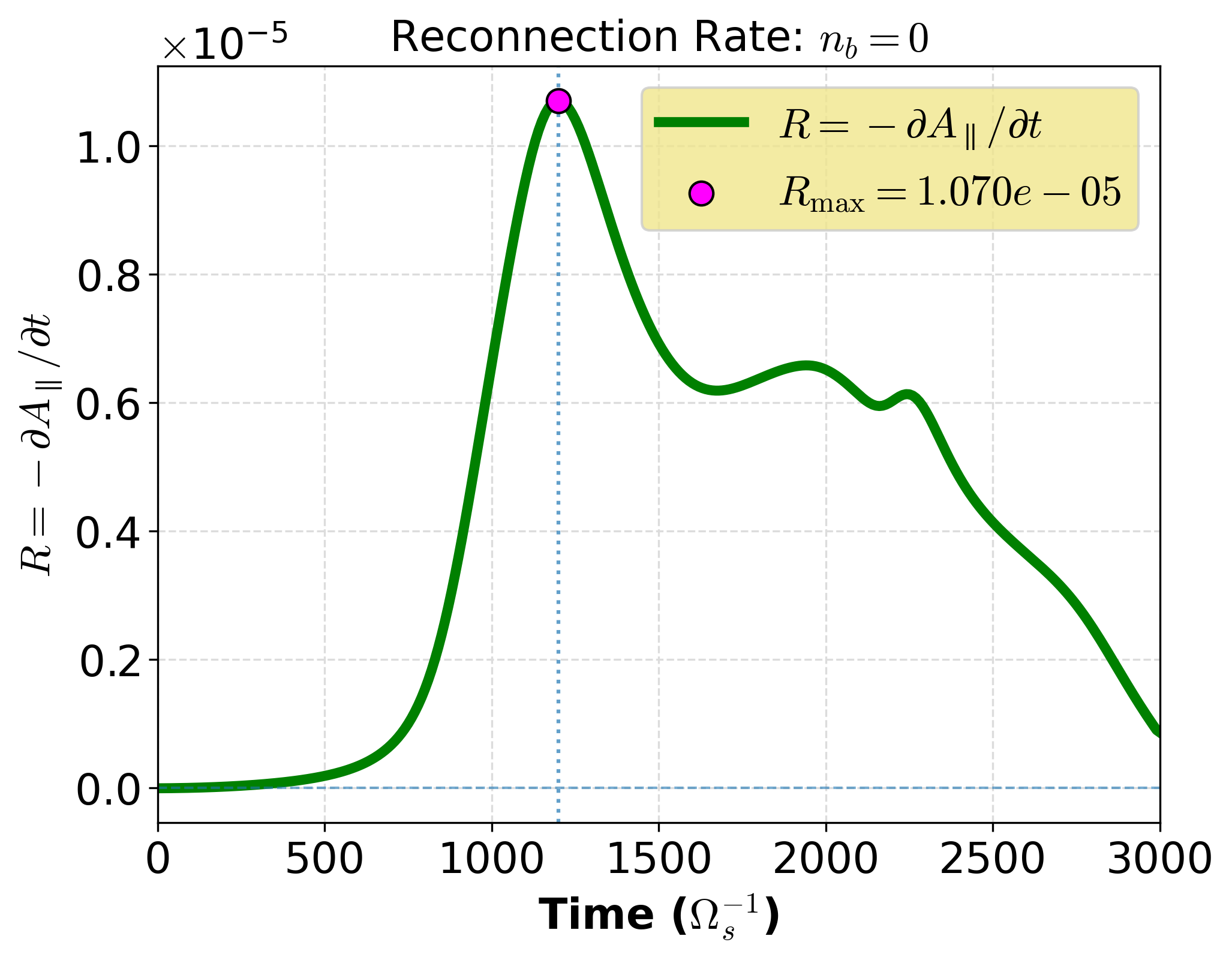}
    \caption{Temporal evolution of the normalized reconnection rate at the X-point for the flat-density case ($n_b=0$). The reconnection rate increases rapidly during filament approach, reaches a maximum of $R_{\max}=1.07\times10^{-5}$ at $t\simeq1200\,\Omega_s^{-1}$, and subsequently decreases as the current sheet relaxes during coalescence.}
    \label{fig:reconnection_rate}
\end{figure}

We next examine the nonlinear reconnection dynamics. Figure~\ref{fig:reconnection_rate} presents the temporal evolution of the normalized reconnection rate, $R=-\partial A_{\parallel}/\partial t$, evaluated at the X-point. During the initial stage, the reconnection rate remains close to zero, indicating that the two filaments evolve independently before a current sheet is formed. As the Lorentz force drives the filaments together, the reconnection rate increases rapidly and reaches its first maximum at approximately $t\approx1200~\Omega_s^{-1}$, corresponding to the formation of a thin current sheet and the onset of fast magnetic reconnection. Following this peak, the reconnection rate decreases as magnetic flux is transferred across the reconnection layer and the current sheet begins to relax.

Interestingly, a second, weaker reconnection peak appears near $t\approx2100~\Omega_s^{-1}$. This secondary peak is associated with the oscillatory relaxation (sloshing) of the merged magnetic structure, during which the reconnecting current sheet undergoes repeated compression and expansion before reaching equilibrium. Similar multiple reconnection peaks have been reported in classical magnetic island coalescence studies and are generally attributed to repeated magnetic flux pile-up during the nonlinear evolution. The temporal evolution of the reconnection rate therefore demonstrates that current filament merging proceeds through the same nonlinear reconnection dynamics as the magnetic island coalescence problem, providing a second independent verification of the correspondence established from the magnetic topology evolution.

\begin{figure}
    \centering
    \includegraphics[width=0.99\linewidth]{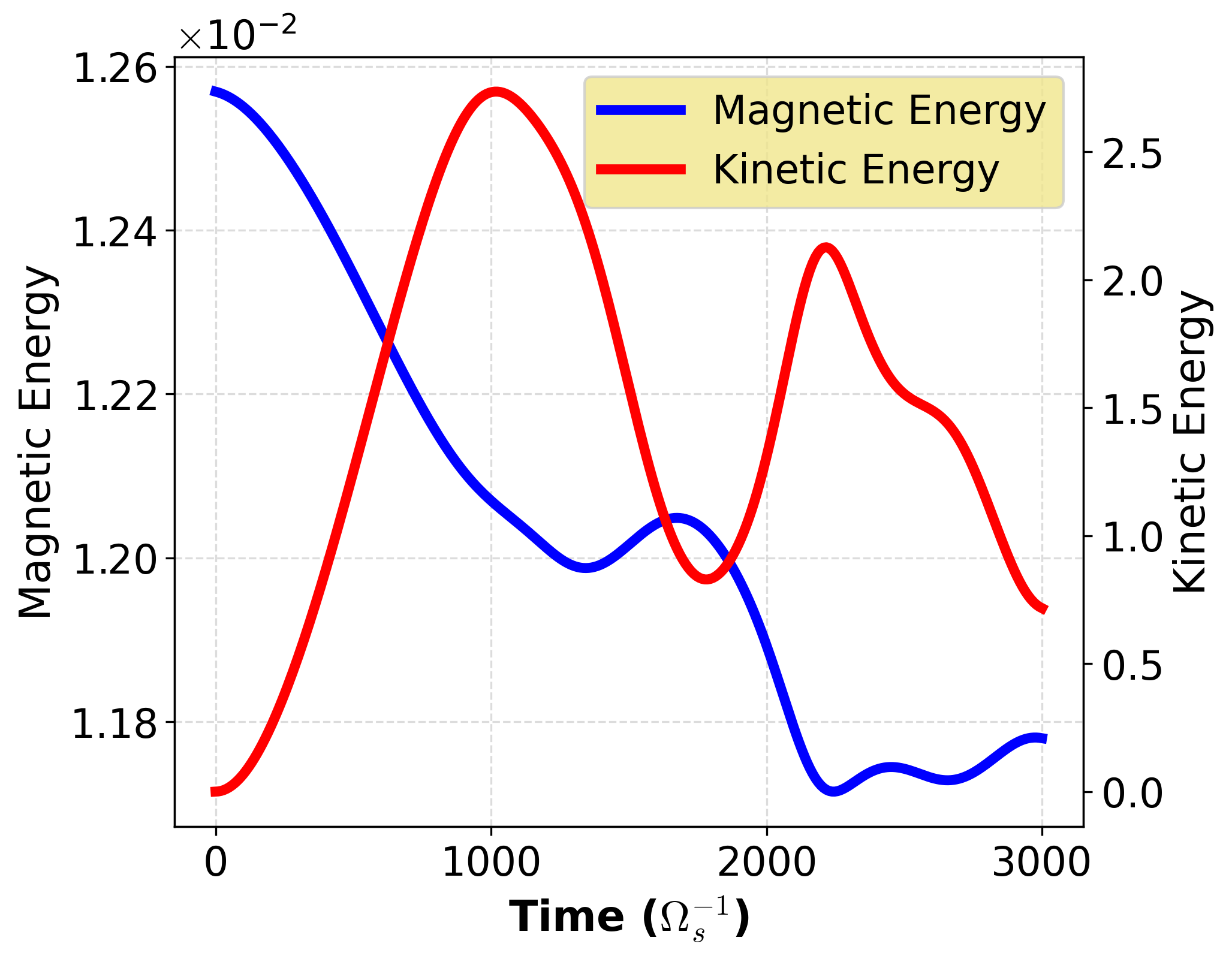}
    \caption{Temporal evolution of the magnetic and kinetic energies during filament coalescence. The decrease in magnetic energy is accompanied by a corresponding increase in kinetic energy, indicating the conversion of magnetic energy into plasma kinetic energy during the merging process.}
    \label{fig:energy}
\end{figure}

To further verify that the coalescence process is governed by magnetic reconnection, we examine the evolution of the magnetic and kinetic energies during the interaction. Figure~\ref{fig:energy} shows the temporal evolution of the total magnetic energy and kinetic energy. Initially, the energy is predominantly stored in the magnetic field associated with the two current filaments, while the kinetic energy remains negligible since the plasma is initially at rest. As the filaments approach each other, the magnetic energy decreases continuously, whereas the kinetic energy increases rapidly, reaching its first maximum at approximately $t\approx1000~\Omega_s^{-1}$. This energy exchange coincides with the formation of the thin current sheet and the onset of rapid magnetic reconnection, indicating an efficient conversion of magnetic energy into plasma kinetic energy. After the first reconnection event, the kinetic energy decreases as the reconnecting current sheet relaxes. A second, weaker enhancement of the kinetic energy is observed around $t\approx2100~\Omega_s^{-1}$, accompanied by a corresponding decrease in magnetic energy. This secondary energy conversion is associated with the sloshing motion of the merged magnetic structure. The anti-correlated evolution of the magnetic and kinetic energies demonstrates that magnetic reconnection is the primary mechanism responsible for the energy release during current filament coalescence, consistent with the energy conversion process observed in magnetic island coalescence studies.

\begin{figure}
    \centering
    \includegraphics[width=0.99\linewidth]{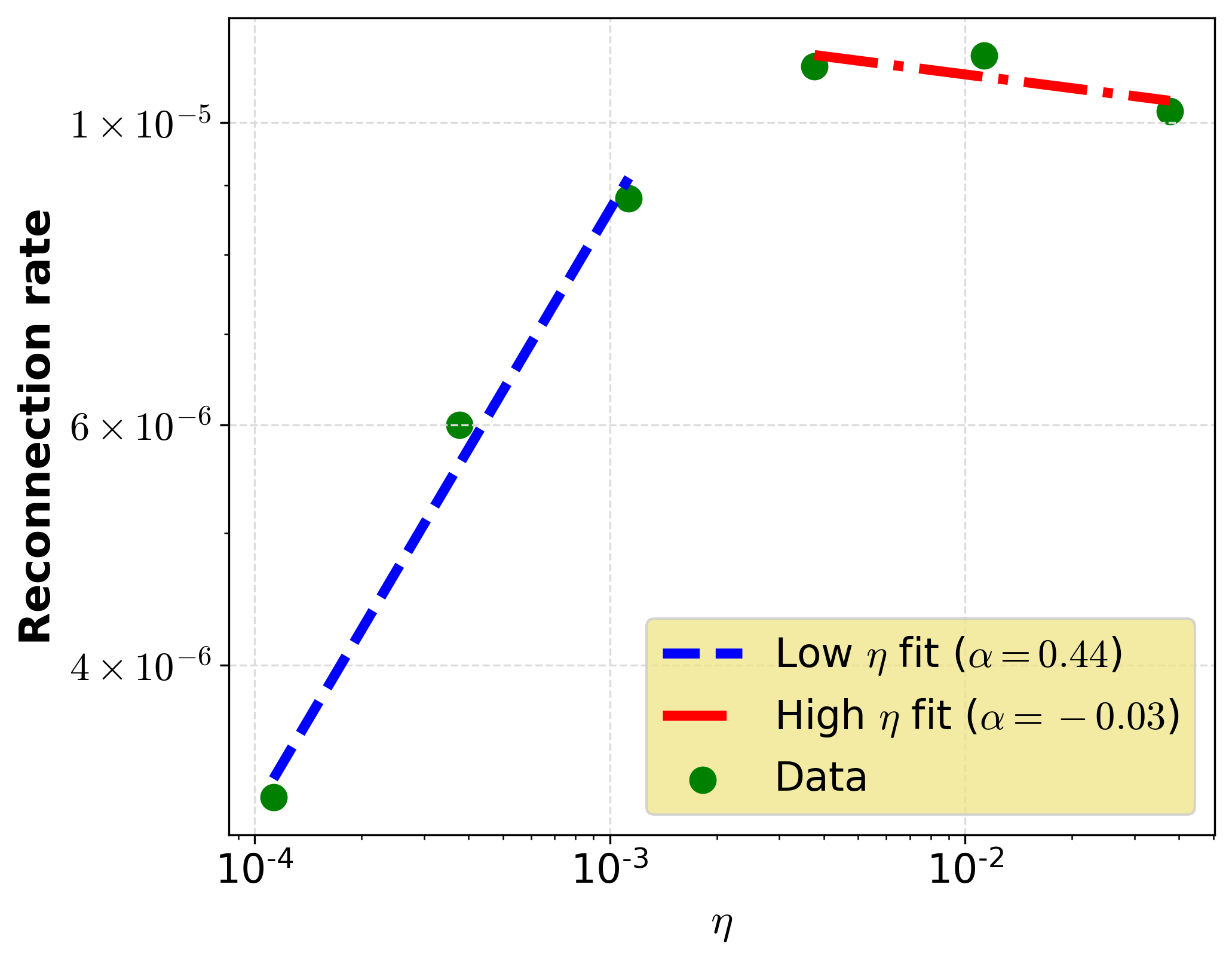}
    \caption{Peak reconnection rate as a function of resistivity $\eta$ in the flat-density limit ($n_b=0$). The low-$\eta$ regime follows a power-law scaling $R_{\rm peak}\propto\eta^{0.44}$, while the reconnection rate becomes $\propto \eta^{-0.03}$ at higher resistivity.}
    \label{fig:rate_scalling}
\end{figure}

With the correspondence between current-filament merging and magnetic-island coalescence established, we next investigate the resistivity dependence of the peak reconnection rate. Figure~\ref{fig:rate_scalling} presents the log-log variation of the maximum reconnection rate with resistivity. The results exhibit two distinct scaling regimes. In the low-resistivity regime, the peak reconnection rate follows the scaling
\begin{equation}
R_{\rm peak}\propto\eta^{0.44},
\end{equation}
which is close to the classical Sweet-Parker prediction, $R_{\rm peak}\propto\eta^{1/2}$, indicating that the reconnection process is controlled primarily by resistive diffusion within the current sheet. As the resistivity increases, the scaling changes to
\begin{equation}
R_{\rm peak}\propto\eta^{0.03},
\end{equation}
demonstrating that the reconnection rate becomes nearly independent of resistivity. This weak dependence suggests a transition from a resistively limited regime to one dominated by the global electromagnetic interaction of the current-carrying filaments. The observed two-regime scaling is qualitatively consistent with previous magnetic island coalescence studies \cite{D_A_Knoll_Pop_2006} and provides strong quantitative evidence that the present filament-merging system follows the same reconnection physics.

\begin{figure}
    \centering
    \includegraphics[width=0.99\linewidth]{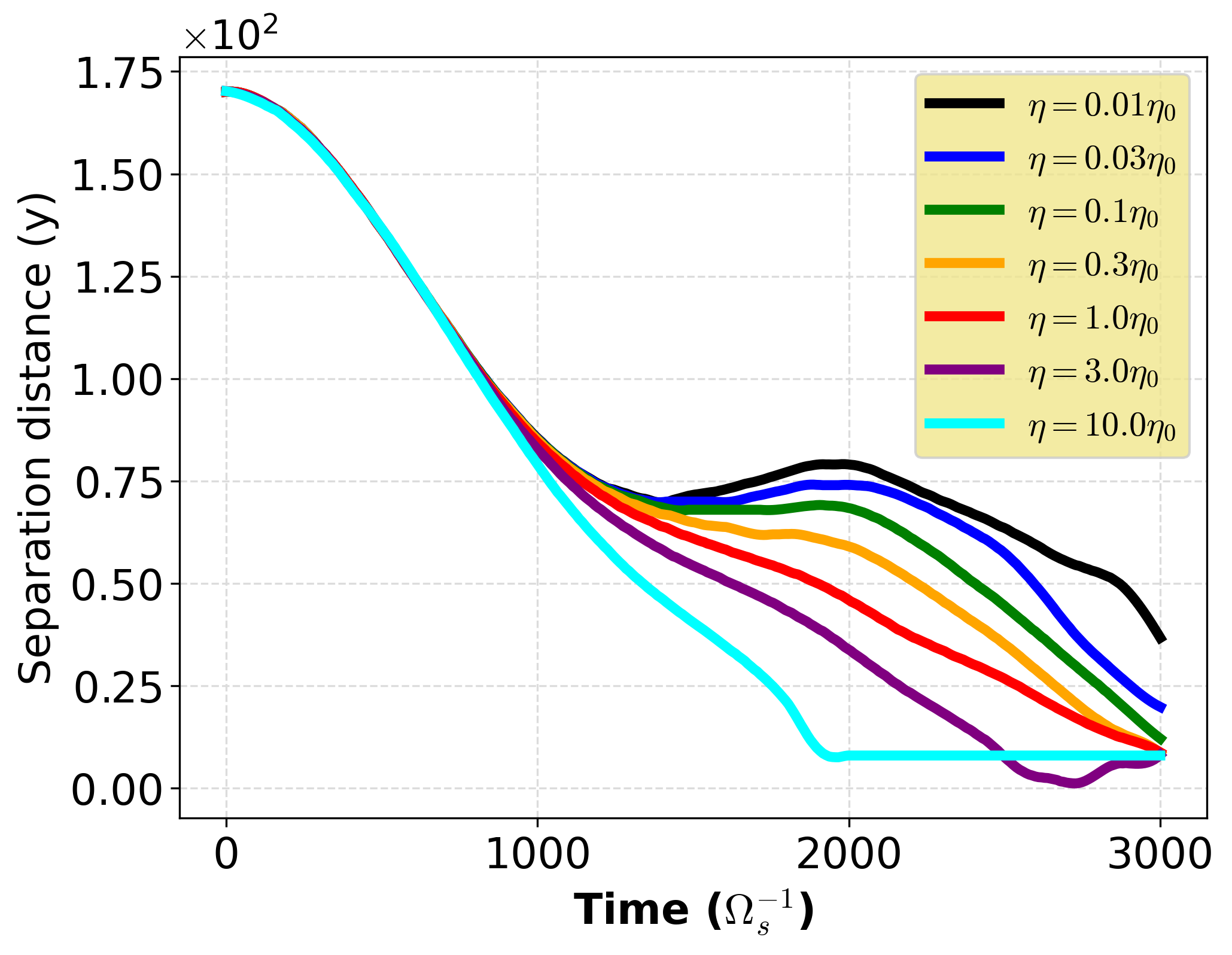}
    \caption{Temporal evolution of the separation distance between the O-points of the two current-carrying filaments for different resistivities. The filaments approach each other for all cases, while the coalescence dynamics and final separation depend strongly on the resistivity.}
    \label{fig:sep_eta}
\end{figure}

Finally, the global coalescence dynamics are quantified by measuring the temporal evolution of the separation distance between the O-points of the two current filaments. Figure~\ref{fig:sep_eta} shows the separation distance for different values of resistivity. During the initial stage, all cases follow nearly identical trajectories, indicating that the early attraction is governed primarily by the Lorentz force and is only weakly influenced by resistivity. As the filaments approach each other and the current sheet forms, the evolution becomes increasingly resistivity-dependent. For low resistivity ($\eta\lesssim0.1$), the separation distance exhibits pronounced oscillations before eventually decreasing to zero. These oscillations correspond to the sloshing motion of the reconnecting magnetic structures caused by repeated compression and rebound of the current sheet. In contrast, increasing the resistivity suppresses the sloshing behavior, resulting in a more monotonic decrease of the separation distance and earlier filament coalescence due to enhanced magnetic diffusion. Such oscillatory and monotonic coalescence regimes have also been reported in classical magnetic island coalescence studies \cite{D_A_Knoll_Pop_2006, D_A_Knoll_PRL_2006}. Together with the magnetic topology evolution, linear perturbation, reconnection-rate history, energy conversion, and resistivity scaling presented above, the separation-distance evolution provides a comprehensive demonstration that the merging of current-carrying filaments in the flat-density limit follows the same nonlinear dynamics as the classical magnetic island coalescence problem.

\subsection{Density Perturbation Effects on Magnetic Reconnection}

\begin{figure*}
    \centering
    \includegraphics[width=\linewidth]{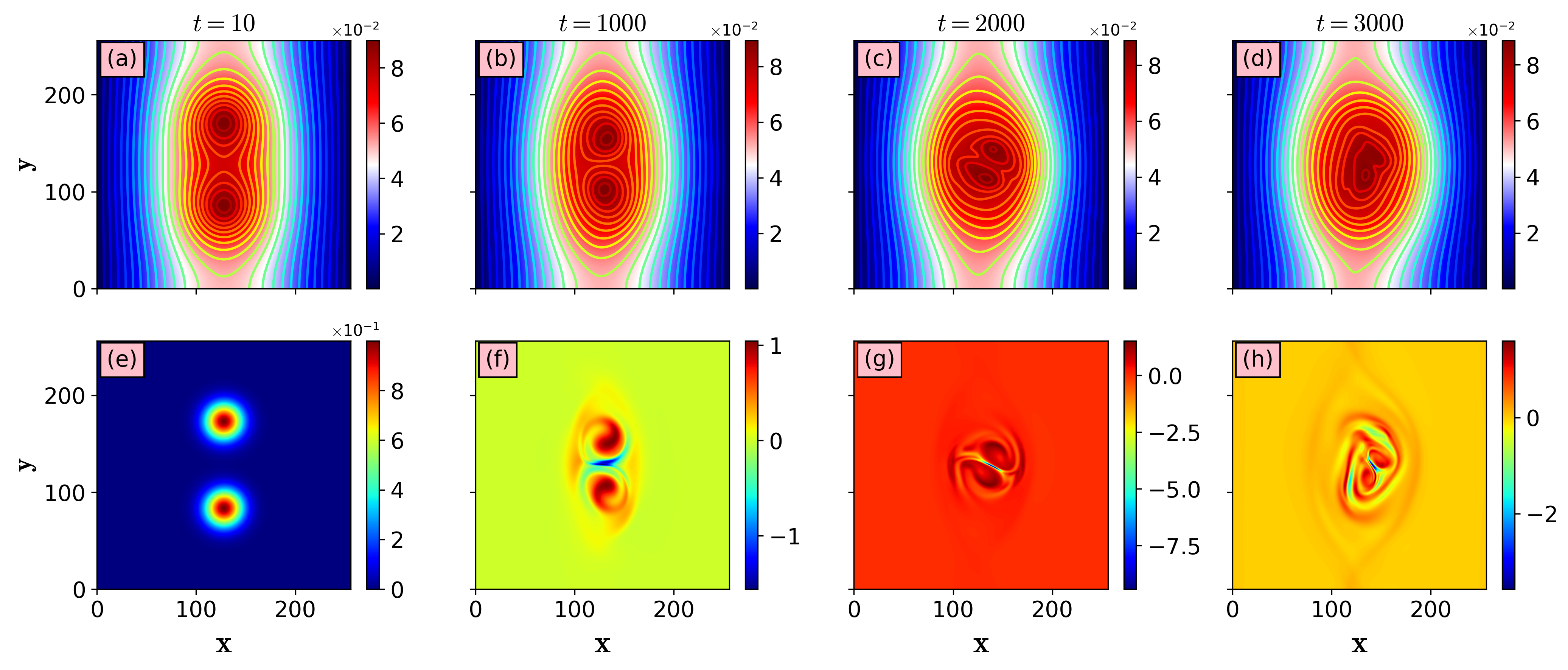}
    \caption{
    Temporal evolution of the magnetic and parallel-current structures for $n_b=1$. The top row [(a)-(d)] shows the distribution of $A_{\parallel}$ (color and contour lines), while the bottom row [(e)-(h)] shows $J_{\parallel}$ at $t=10$, $1000$, $2000$, and $3000\,\Omega_s^{-1}$, respectively. The initially separated magnetic structures approach each other, develop a localized current-sheet region, and subsequently evolve into a strongly deformed magnetic configuration accompanied by complex parallel current structures.}
    \label{fig:Ap_Jp_evolution}
\end{figure*}

Having established the reconnection-driven coalescence of the two current-carrying filaments in the flat-density limit, we next examine the magnetic reconnection process in the presence of a finite density perturbation. The blob-like density perturbation introduces finite density and pressure gradients, thereby modifying the plasma motion and magnetic evolution during filament interaction. Figure~\ref{fig:Ap_Jp_evolution} shows the spatial distributions of $A_{\parallel}$ (top row) and $J_{\parallel}$ (bottom row) at four representative times, $t=10$, $1000$, $2000$, and $3000\,\Omega_s^{-1}$, for the case $n_b=1$. At the initial stage, $t=10\,\Omega_s^{-1}$, the two filaments are clearly separated in the poloidal direction, as reflected by the two distinct extrema of $A_{\parallel}$ in Fig.~\ref{fig:Ap_Jp_evolution}(a). The corresponding parallel-current distribution consists of two well-localized current concentrations [Fig.~\ref{fig:Ap_Jp_evolution}(e)]. As the filaments approach one another, the magnetic structures become increasingly distorted. At $t=1000\,\Omega_s^{-1}$, the two magnetic structures are still identifiable, but their contours become strongly deformed, and the associated current distributions develop pronounced opposite-sign structures around the interaction region [Fig.~\ref{fig:Ap_Jp_evolution}(b,f)]. This indicates the development of a localized current sheet between the interacting magnetic structures.

At later times, the two structures undergo substantial nonlinear reorganization. By $t=2000\,\Omega_s^{-1}$, the original two-centre magnetic configuration has evolved into a single, strongly distorted magnetic structure, accompanied by a localized and spatially structured parallel-current distribution [Fig.~\ref{fig:Ap_Jp_evolution}(c,g)]. At $t=3000\,\Omega_s^{-1}$, the magnetic configuration exhibits a pronounced asymmetric, swirling structure, while the current distribution develops an extended, highly deformed pattern [Fig.~\ref{fig:Ap_Jp_evolution}(d,h)]. Thus, the initially separated current-carrying filaments do not simply coalesce into a stationary configuration; instead, the magnetic and current structures continue to evolve after the initial interaction.
The evolution shown in Fig.~\ref{fig:Ap_Jp_evolution} provides the spatial picture underlying the subsequent reconnection dynamics. In particular, the formation and deformation of the localized current-sheet region indicate that magnetic-flux transfer occurs through a dynamically evolving reconnection region rather than at a fixed spatial location.


\begin{figure}
    \centering
    \includegraphics[width=0.95\linewidth]{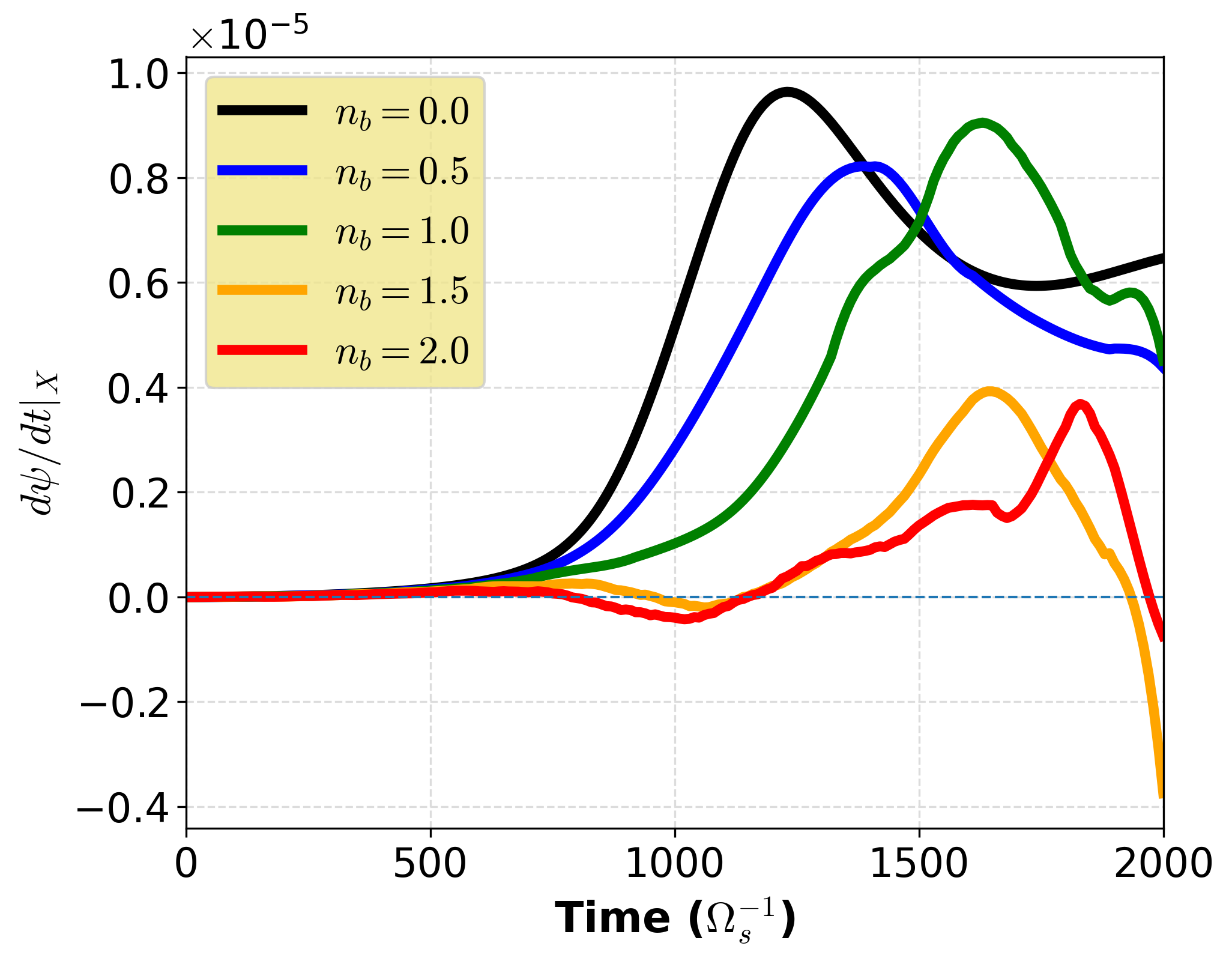}
    \caption{
    Temporal evolution of the reconnection rate, $R=-\left.\partial A_{\parallel}/\partial t\right|_{X}$, evaluated at the instantaneous X-point for different values of the density perturbation amplitude, $n_b$. The reconnection rate for $n_b=0$, $0.5$, and $1.0$ reaches comparable peak values, while the peak magnitude decreases for $n_b=1.5$ and $2.0$. The time of the maximum reconnection rate also shifts progressively toward later times with increasing $n_b$, indicating a delayed and suppressed reconnection response for stronger density perturbations.}
    \label{fig:reconnection_rate_nb}
\end{figure}

Consequently, in the following analysis, the reconnection rate is evaluated at the instantaneous X-point of the magnetic configuration. The reconnection rate is evaluated at the instantaneous X-point as
\begin{equation}
R(t) =
-\left.\frac{\partial A_{\parallel}}{\partial t}\right|_{X},
\end{equation}
following the conventional characterization of magnetic reconnection through the evolution of the reconnecting magnetic flux \cite{Yamada_2010}. Figure~\ref{fig:reconnection_rate_nb} shows the temporal evolution of the reconnection rate for different values of the density perturbation amplitude, $n_b$. For $n_b=0$, the reconnection rate increases rapidly after the initial stage and reaches a maximum of approximately $9.7\times10^{-6}$ at $t\simeq1200\,\Omega_s^{-1}$. As the density perturbation is increased to $n_b=0.5$ and $1.0$, the peak reconnection rate remains of comparable magnitude, although the time at which the maximum is attained shifts progressively toward later times. For stronger density perturbations, $n_b>1$, a pronounced reduction in the peak reconnection rate is observed. The maximum reconnection rate decreases to approximately $3.9\times10^{-6}$ and $3.7\times10^{-6}$ for $n_b=1.5$ and $2.0$, respectively. At the same time, the corresponding peak is progressively delayed, occurring at approximately $t\simeq1650\,\Omega_s^{-1}$ for $n_b=1.5$ and $t\simeq1830\,\Omega_s^{-1}$ for $n_b=2.0$. These results indicate that the density perturbation affects both the strength and timing of the reconnection process. For moderate density perturbations, $n_b\lesssim1$, the peak reconnection rate remains relatively insensitive to $n_b$, while for stronger perturbations the reconnection rate is suppressed and its maximum is shifted to later times.

\begin{figure*}
    \centering
    \includegraphics[width=0.45\linewidth]{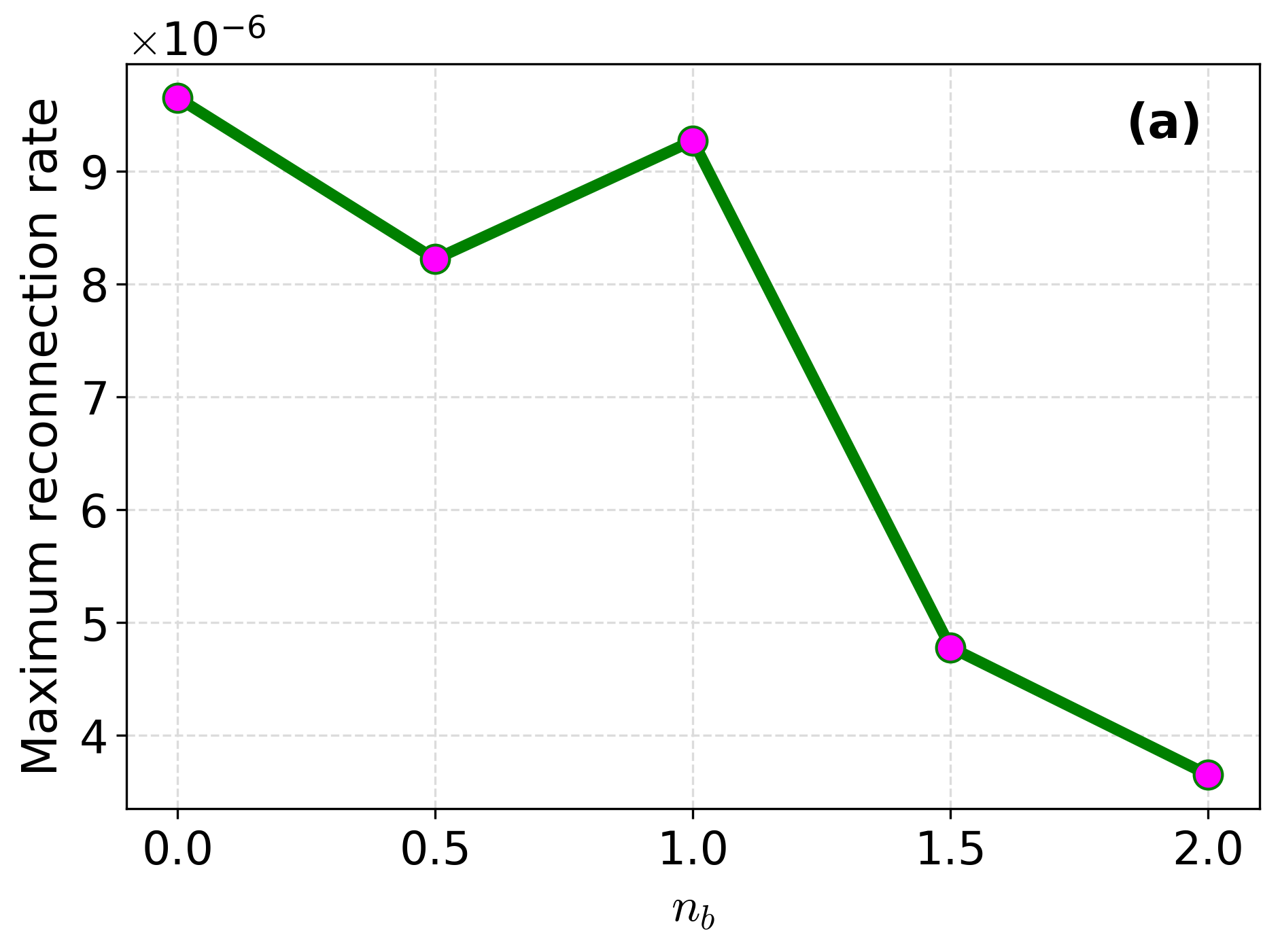}
    \includegraphics[width=0.45\linewidth]{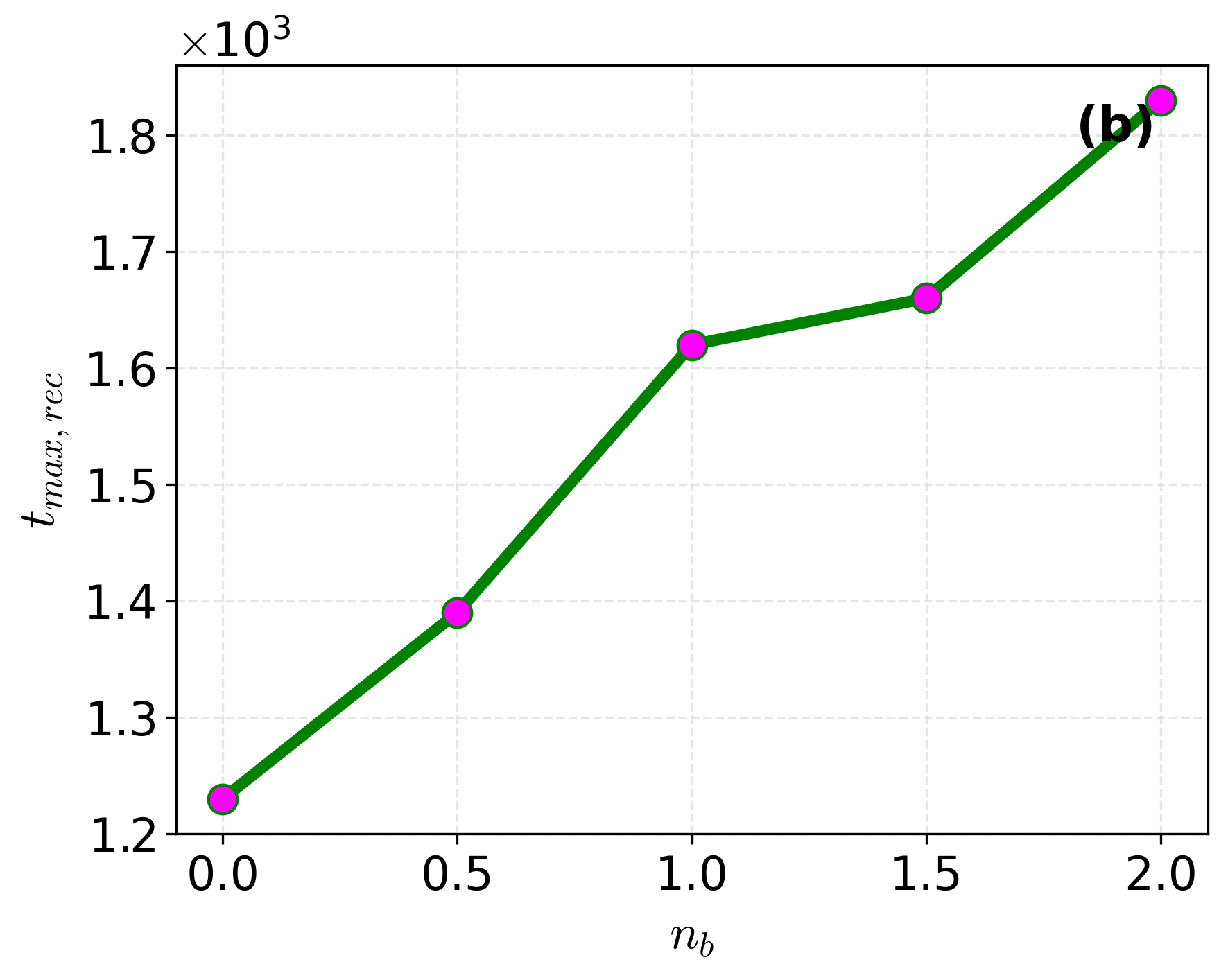}
    \caption{Dependence of the reconnection dynamics on the density perturbation amplitude $n_b$. (a) Maximum reconnection rate, $R_{\rm peak}$, evaluated at the instantaneous X-point. (b) Time at which the maximum reconnection rate is attained, $t_{\max,\mathrm{rec}}$.}
    \label{fig:reconnection_summary}
\end{figure*}

To quantify the dependence of the reconnection dynamics on the density perturbation, we extract the maximum reconnection rate and the corresponding time from the temporal evolution shown in Fig.~\ref{fig:reconnection_summary}. As shown in Fig.~\ref{fig:reconnection_summary} (left), the maximum reconnection rate remains approximately comparable for $n_b\leq1$, indicating that moderate density perturbations do not substantially modify the peak reconnection strength. However, for $n_b>1$, $R_{\max}$ decreases markedly with increasing $n_b$. In contrast, the time corresponding to the maximum reconnection rate increases systematically with $n_b$, as shown in Fig.~\ref{fig:reconnection_summary} (right). Thus, increasing the density perturbation progressively delays the occurrence of the maximum reconnection rate and, for sufficiently large $n_b$, also reduces its magnitude.


\begin{figure}
    \centering
    \includegraphics[width=0.95\linewidth]{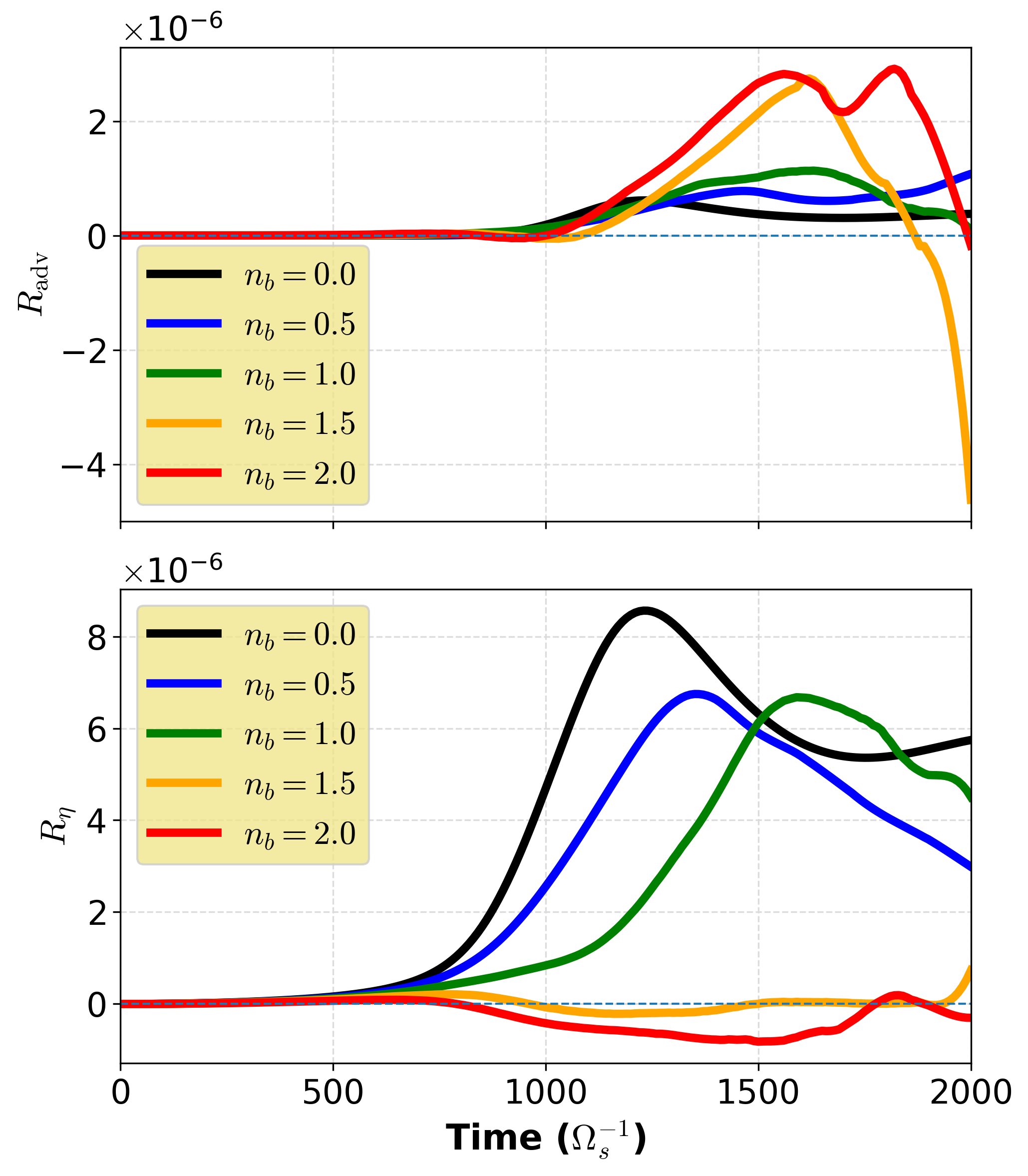}
    \caption{Temporal evolution of the two contributions to the parallel induction equation evaluated at the instantaneous X-point for different density perturbation amplitudes $n_b$. The upper panel shows the advective contribution $R_{\mathrm{adv}}=-[A_{\parallel},\ln n]$, while the lower panel shows the resistive contribution $R_{\eta}=\eta\nabla_{\perp}^{2}A_{\parallel}$. For $n_b\leq1$, the resistive contribution dominates during the main reconnection phase. For larger density perturbations, $n_b=1.5$ and $2.0$, the resistive contribution is substantially reduced, while the advective contribution becomes progressively stronger, indicating a transition toward density-gradient control of the magnetic-potential evolution.}
    \label{fig:induction_terms}
\end{figure}

To identify the physical origin of the dependence of the reconnection rate on the density perturbation, we decompose the parallel induction equation into its advective and resistive contributions. The two contributions evaluated at the instantaneous X-point are shown in Fig.~\ref{fig:induction_terms}. The advective contribution is defined as
\begin{equation}
R_{\mathrm{adv}}=-\left[\!A_{\parallel},\ln n\!\right],
\end{equation}
whereas the resistive contribution is given by
\begin{equation}
R_{\eta}=\eta\nabla_{\perp}^{2}A_{\parallel}.
\end{equation}

For weak density perturbations, $n_b\leq1$, the resistive contribution provides the dominant positive contribution during the main reconnection phase. In particular, $R_{\eta}$ increases strongly as the two filaments approach each other and reaches values of several $10^{-6}$, whereas the advective contribution remains comparatively small. The temporal evolution of $R_{\eta}$ also follows the evolution of the reconnection rate shown in Fig.~\ref{fig:reconnection_rate_nb}, indicating that the reconnection process in this regime is predominantly controlled by the resistive current-sheet contribution. Such a resistive current-sheet evolution is consistent with the established picture of magnetic-island coalescence and reconnection
\cite{biskamp_PRA_1982, D_A_Knoll_Pop_2006}.  A qualitatively different behavior emerges for stronger density perturbations. As $n_b$ is increased to $1.5$ and $2.0$, the resistive contribution is strongly reduced and remains close to zero or becomes slightly negative during a substantial part of the reconnection phase. In contrast, the magnitude of the advective contribution increases significantly with $n_b$. For $n_b=1.5$ and $2.0$, $R_{\mathrm{adv}}$ grows to values of order $10^{-6}$ to $3\times10^{-6}$ and becomes the dominant positive contribution during the later evolution. Thus, the increase in density perturbation does not simply enhance the resistive current-driven contribution; instead, it progressively changes the relative importance of the terms governing the evolution of $A_{\parallel}$.

This behavior provides a physical explanation for the results shown in Fig.~\ref{fig:reconnection_summary}. For $n_b\lesssim1$, the reconnection dynamics remain primarily associated with the resistive term, and the maximum reconnection rate remains relatively unchanged. For larger $n_b$, the increasing density gradients introduce a stronger advective contribution through the $[A_{\parallel},\ln n]$ term. This modifies the evolution of the magnetic potential and shifts the time at which the maximum reconnection response is attained. At the same time, the reduction
of the resistive contribution results in a lower maximum reconnection rate. Therefore, the increasing $n_b$ leads to a transition from a predominantly resistive reconnection regime to a regime in which the density-dependent advective term plays an increasingly important role.


\begin{figure}
    \centering
    \includegraphics[width=0.95\linewidth]{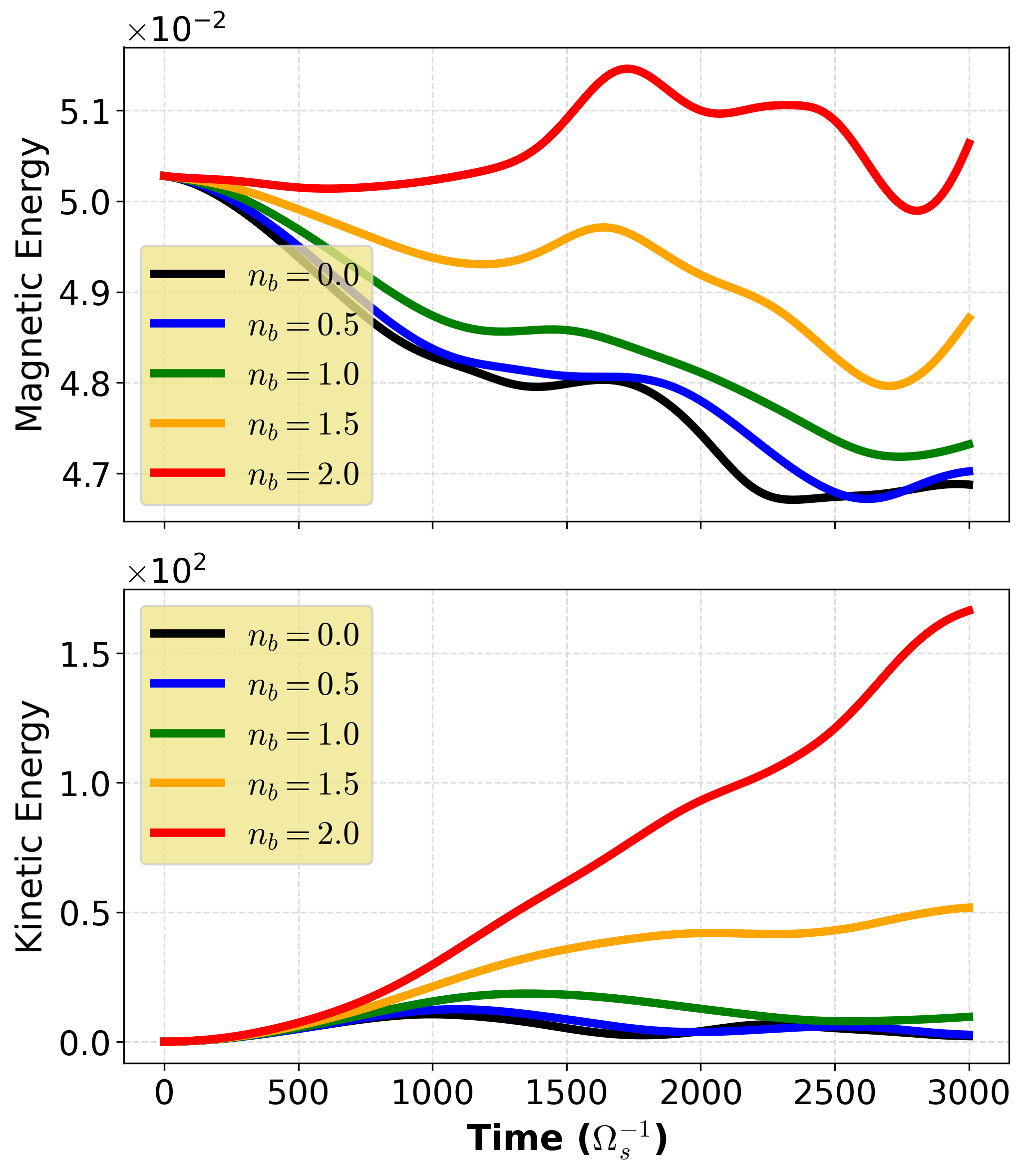}
    \caption{Temporal evolution of the magnetic energy $E_B$ (top) and perpendicular kinetic energy $E_K$ (bottom) for different values of the density perturbation amplitude $n_b$. Increasing $n_b$ produces increasingly pronounced temporal variations in the magnetic energy and a strong enhancement of the kinetic-energy response. The results indicate enhanced redistribution of energy between the magnetic and perpendicular-flow degrees of freedom for larger density perturbations.}
    \label{fig:energy_evolution}
\end{figure}

The evolution of the magnetic and kinetic energies provides a complementary measure of the energy redistribution during the interaction. Figure~\ref{fig:energy_evolution} shows the temporal
evolution of the magnetic energy $E_B$ and the perpendicular kinetic energy $E_K$ for different values of the density perturbation amplitude $n_b$. For the unperturbed case, $n_b=0$, the magnetic energy decreases progressively during the filament interaction, while the kinetic energy initially increases and subsequently remains relatively small. This behavior is consistent with the conversion of magnetic energy associated with the interacting current structures into perpendicular plasma motion. With increasing $n_b$, the energy evolution changes substantially. The magnetic energy remains at a higher level throughout the evolution, and its temporal variation becomes increasingly non-monotonic. In particular, for $n_b=1.5$ and $2.0$, the magnetic energy exhibits pronounced variations, including a temporary increase during the interaction. At the same time, the kinetic energy increases strongly with $n_b$. The enhancement is particularly pronounced for $n_b=2$, for which the kinetic energy continues to grow throughout the simulation and reaches a value much larger than that obtained for the lower-$n_b$ cases.

The simultaneous enhancement of the kinetic-energy response and the stronger temporal modulation of the magnetic energy indicate that the density perturbation substantially modifies the redistribution of energy between the magnetic and flow degrees of freedom. Thus, the increasing $n_b$ does not simply correspond to a stronger monotonic release of magnetic energy; rather, it produces a more dynamically active magnetic-kinetic energy exchange. This enhanced energy redistribution is consistent with the delayed and modified reconnection dynamics observed in Figs.~\ref{fig:reconnection_rate_nb} and~\ref{fig:reconnection_summary}.

\subsection{Density Perturbation Enhancement of Filament Sloshing}


\begin{figure}
    \centering
    \includegraphics[width=0.95\linewidth]{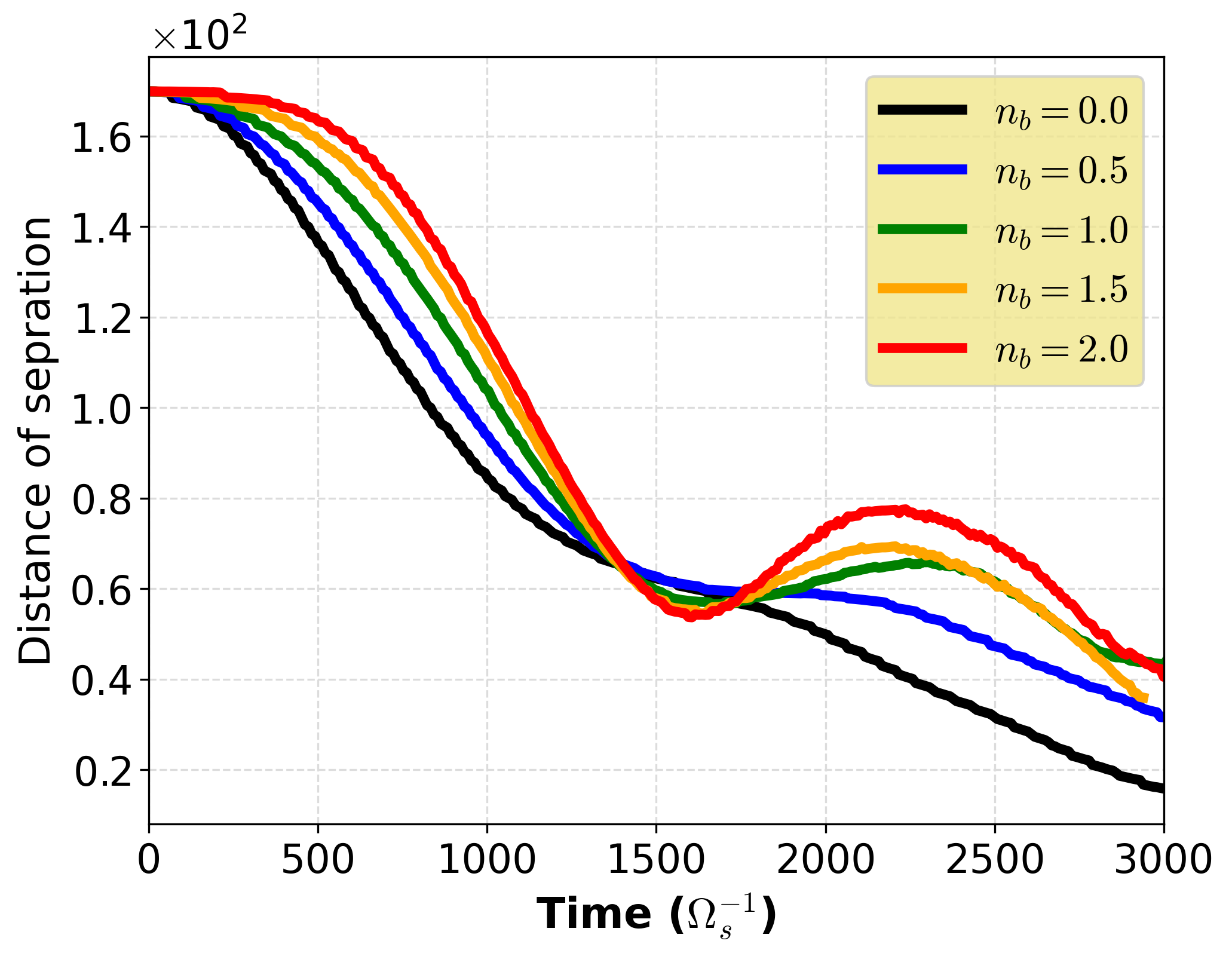}
    \caption{Temporal evolution of the separation distance between the two interacting filaments for different density perturbation amplitudes, $n_b$. All cases start from nearly the same initial separation. During the initial approach phase, the separation decreases for all cases and reaches a first minimum at $t\simeq1500$--$1700\,\Omega_s^{-1}$. For finite $n_b$, the filaments subsequently rebound, with the amplitude of the post-minimum separation increasingly enhanced as $n_b$ is increased. The $n_b=0$ case exhibits predominantly monotonic contraction over the analyzed interval, whereas finite density perturbations produce a clear rebound characteristic of the sloshing motion.}
    \label{fig:separation}
\end{figure}

The effect of the density perturbation on the subsequent filament motion is examined through the evolution of the separation distance between the two filaments, as shown in Fig.~\ref{fig:separation}. All cases start from nearly the same initial separation, allowing the effect of $n_b$ on the subsequent dynamics to be compared directly. During the initial approach phase, the separation decreases for all values of $n_b$. However, the rate of approach depends systematically on the density perturbation. The $n_b=0$ case exhibits the fastest initial decrease in separation and continues to contract after the first minimum, with no pronounced rebound over the simulated time interval. A qualitatively different behavior is observed for finite density
perturbations. The separation reaches a first minimum at approximately $t\simeq 1500$--$1700\,\Omega_s^{-1}$, after which the filaments begin to move apart again. This post-minimum increase in separation becomes progressively stronger with increasing $n_b$. In particular, the rebound is weak for $n_b=0.5$, becomes more pronounced for $n_b=1$, and is further enhanced for $n_b=1.5$ and $2.0$. The largest rebound is obtained for $n_b=2$, for which the separation increases to nearly $80$ before subsequently decreasing again.

The sloshing of interacting magnetic islands is a well-established feature of magnetic-island coalescence, particularly in the low-resistivity (high-Lundquist-number) regime. Previous resistive-MHD studies have shown that the coalescence can stall and the islands can rebound because the dynamically thinning current sheet produces a strong opposing magnetic-pressure gradient
\cite{D_A_Knoll_Pop_2006, D_A_Knoll_PRL_2006}. Similar oscillatory or sloshing behavior has also been reported in magnetic-flux-rope and island-coalescence studies. The present results demonstrate an additional mechanism for enhancing this behavior. Here, the resistivity is kept fixed, while the amplitude of the density perturbation is varied. The systematic increase in the post-minimum rebound with $n_b$ therefore indicates that the density perturbation provides an additional contribution to the restoring dynamics responsible for the sloshing motion.


To quantify the rebound, we define the first-cycle sloshing amplitude as
\begin{equation}
A_{\rm slosh}=d_{\rm reb}-d_{\rm min},\label{eq:Aslosh}
\end{equation}
where $d_{\rm min}$ is the first minimum of the separation distance and $d_{\rm reb}$ is the first local maximum following this minimum. For $n_b=0$ and $0.5$, no clearly resolved post-minimum rebound is observed within the analyzed interval, whereas a finite and progressively larger $A_{\rm slosh}$ is obtained for $n_b\geq1$. Thus, the density perturbation does not merely modify the initial approach of the filaments but enhances the subsequent sloshing response.


\begin{figure}
    \centering
    \includegraphics[width=0.95\linewidth]{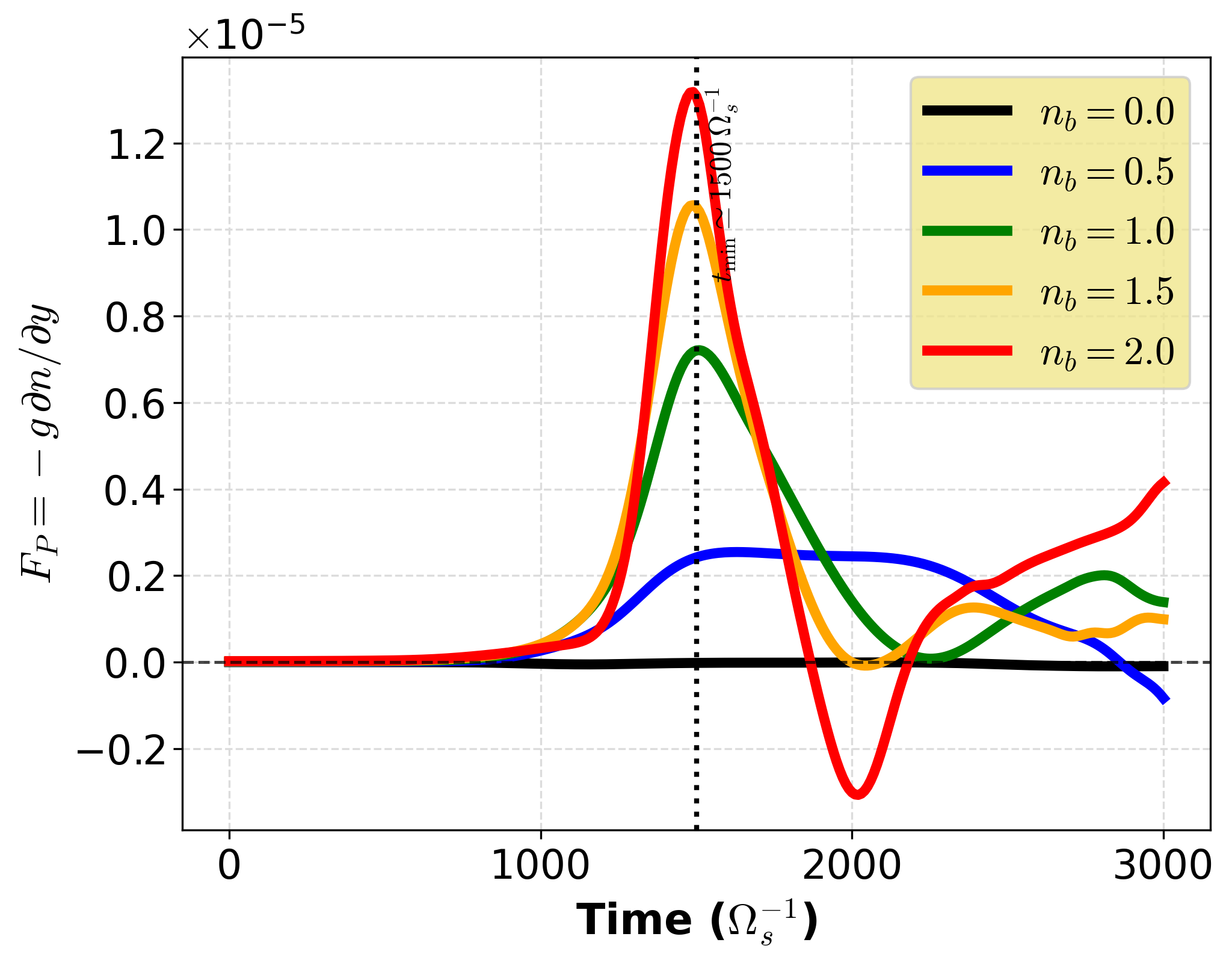}
    \caption{Temporal evolution of the maximum pressure-gradient force, $F_P=-g\,\partial n/\partial y$, for different density perturbation amplitudes $n_b$. The vertical dotted line marks  $t=1500\,\Omega_s^{-1}$, approximately corresponding to the first minimum of the filament separation. The pressure force increases strongly with $n_b$ and reaches its largest value near the maximum-compression stage. For the larger density perturbations, $n_b=1.5$ and $2.0$, the force subsequently changes sign during the rebound phase, indicating a reversal of the pressure-gradient response as the density structure reorganizes.}
    \label{fig:pressure_force_time}
\end{figure}

The origin of the enhanced rebound can be further examined from the density-gradient force appearing in the vorticity dynamics. The pressure-gradient force is defined as
\begin{equation}
F_P=-g\frac{\partial n}{\partial y}. \label{eq:Fpressure}
\end{equation}
Figure~\ref{fig:pressure_force_time} shows the temporal evolution of the pressure-gradient force for different values of $n_b$. For $n_b=0$, the pressure contribution is negligibly small, whereas a finite density perturbation produces an increasingly strong pressure response. The maximum pressure force increases systematically with $n_b$ and becomes particularly pronounced for $n_b\geq1$.

An important feature is that the pressure force reaches its largest magnitude near $t\simeq1500\,\Omega_s^{-1}$, close to the time at which the separation distance reaches its first minimum. This temporal correlation indicates that the strongest density-gradient response develops at the maximum-compression stage of the filament interaction. For the larger density perturbations, the subsequent reversal of the pressure force is also more pronounced, indicating a change in the direction of the pressure-gradient response during the rebound phase.

\begin{figure*}
    \centering
    \includegraphics[width=\linewidth]{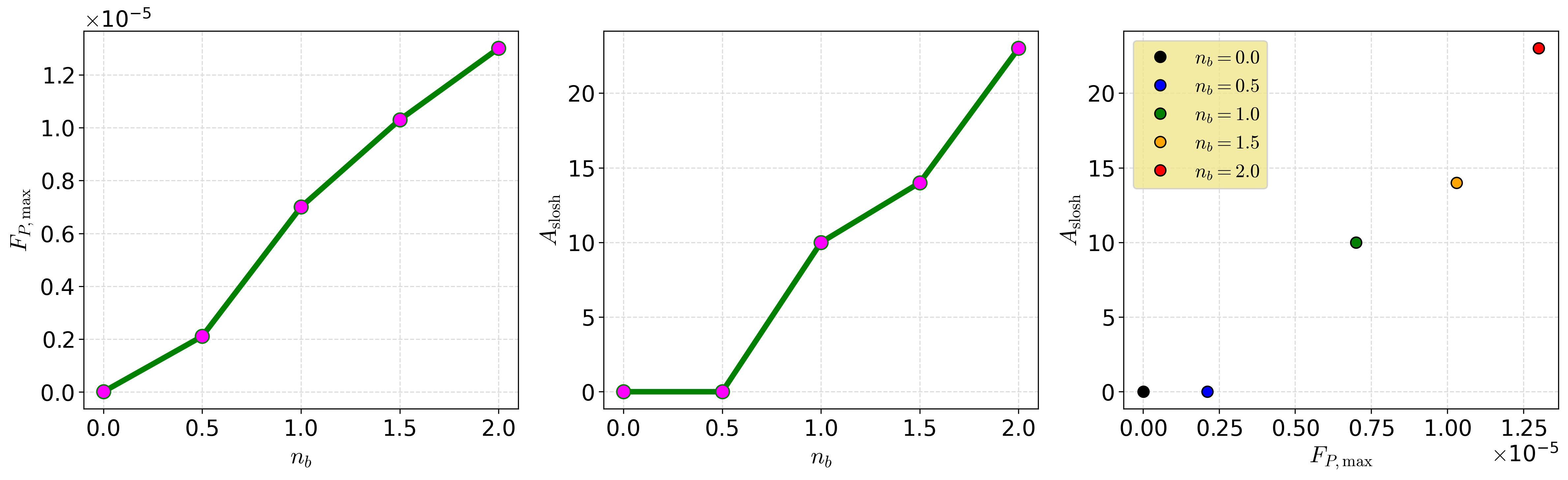}
    \caption{ Quantification of the pressure-force contribution to the sloshing dynamics. (a) Maximum pressure-gradient force, $F_{P,\max}=\max| -g\,\partial n/\partial y |$, as a function of the density perturbation amplitude $n_b$. (b) First-cycle sloshing amplitude, $A_{\rm slosh}=d_{\rm reb}-d_{\rm min}$, as a function of $n_b$. (c) Direct correlation between $A_{\rm slosh}$ and $F_{P,\max}$, with the color indicating $n_b$. The increasing pressure force for larger $n_b$ is accompanied by a systematic enhancement of the post-compression rebound.}
    \label{fig:pressure_slosh}
\end{figure*}

The dependence of the pressure force and sloshing amplitude on the density perturbation is summarized in Fig.~\ref{fig:pressure_slosh}. The maximum pressure force,
\begin{equation}
F_{P,\max}= \max\left| -g\frac{\partial n}{\partial y}\right|,
\end{equation}
increases systematically with $n_b$, while the sloshing amplitude also increases for $n_b\geq1$. Moreover, the direct correlation between $F_{P,\max}$ and $A_{\rm slosh}$ shows that cases with a
stronger pressure-gradient response exhibit a larger post-compression rebound. These results provide evidence that the density perturbation enhances the sloshing dynamics through the pressure-gradient force generated during filament compression.

The observed behavior can therefore be understood as follows. As the two filaments approach each other, the finite density perturbation produces increasingly strong density gradients in the interaction region. The corresponding pressure-gradient force opposes further compression and becomes strongest near the minimum separation. When the compression is sufficiently strong, this restoring response reverses the relative motion of the filaments, producing the observed post-minimum rebound. Increasing $n_b$ strengthens this response and thereby increases the sloshing amplitude. Hence, in addition to the well-established resistivity and current-sheet effects associated with magnetic-island sloshing \cite{biskamp_PRA_1982, D_A_Knoll_PRL_2006}, the present
results identify the density perturbation as an additional control parameter for the strength of the sloshing dynamics.
\section{Discussion and Conclusions}


In this work, we have investigated the nonlinear coalescence dynamics of two parallel current-carrying plasma filaments using a three-dimensional electromagnetic fluid model. In the flat-density limit, $n_b=0$, the filament dynamics exhibit a close analogy with the classical coalescence of neighboring magnetic islands, where parallel currents produce an attractive interaction and drive the structures toward coalescence through magnetic reconnection \cite{Coalescence_instability_finn_kaw, biskamp_PRA_1982}. This establishes magnetic-island coalescence as a useful framework for understanding the merging
dynamics of current-carrying filaments.

We then examined the effect of finite density perturbations on the reconnection dynamics. The reconnection rate evaluated at the instantaneous X-point remains approximately comparable for $n_b\lesssim1$, although the time of its maximum increases with $n_b$. For stronger density perturbations, $n_b>1$, the maximum reconnection rate decreases while its occurrence is further delayed. Analysis of the parallel induction equation shows that this behavior is associated with a change in the relative importance of the different contributions. The resistive contribution dominates for weak density perturbations, consistent with the conventional resistive reconnection \cite{Yamada_2010}, whereas the density-dependent $[A_{\parallel},\ln n]$ contribution becomes increasingly important for larger $n_b$. The corresponding magnetic and kinetic energy evolution also indicates enhanced energy redistribution between the magnetic and perpendicular-flow components as the density perturbation increases.

Finally, we investigated the subsequent filament motion through the evolution of the separation distance. While the flat-density case exhibits predominantly monotonic coalescence, finite density
perturbations produce a post-compression rebound that becomes progressively stronger with increasing $n_b$. Sloshing during magnetic island coalescence has previously been reported, particularly in the low-resistivity regime, and has been associated with the nonlinear evolution of the reconnecting current sheet and the resulting pressure response \cite{biskamp_PRA_1982, D_A_Knoll_Pop_2006}. In the present system, however, the resistivity is kept fixed while the density perturbation is varied. We find that the maximum density-gradient pressure force increases systematically with $n_b$ and becomes strongest near the time of minimum filament separation. Furthermore, the positive correlation between the pressure force and the sloshing amplitude demonstrates that stronger density-gradient forces are associated with a stronger post-compression rebound. Thus, finite density perturbations provide an additional mechanism for enhancing the sloshing dynamics.

Overall, the results establish a unified picture in which the flat-density limit recovers the classical magnetic-island-like coalescence of current-carrying filaments, while finite density perturbations introduce additional density-gradient effects that modify both the reconnection process and the subsequent filament motion. The density perturbation therefore provides an additional control parameter for the nonlinear evolution, reconnection, and sloshing of current-carrying filaments in the edge region of tokamak plasma.


\begin{acknowledgments}


The simulations were carried out on the Antya cluster at the Institute for Plasma Research (IPR) as part of the first author's Ph.D. research. A. Sen gratefully acknowledges INSA for the Honorary Scientist position.
 
\end{acknowledgments}

\bibliographystyle{unsrt}
\bibliography{citation}

\end{document}